\documentclass[aps,floatfix,pre,notitlepage,superscriptaddress,twocolumn]{revtex4-2}
\usepackage{graphicx}
\usepackage[colorlinks=true,allcolors=blue]{hyperref}
\usepackage{mathtools}
\usepackage{physics}
\usepackage{enumitem}
\usepackage{amsmath,amssymb}
\usepackage[dvipsnames]{xcolor}

\newcommand*{\kT}{k_\text{B}T}
\newcommand*{\fa}{f_\text{a}}
\newcommand*{\noiseT}{\bar{\eta}^\text{T}}
\newcommand*{\noiseR}{\eta^\text{R}}
\newcommand*{\Rv}{R_\text{ves}}
\newcommand*{\BM}{\kappa_\text{B}}
\newcommand*{\sm}{\kappa_\text{S}}

\newcommand*{\ULJ}{U_\text{WCA}}
\newcommand*{\Frigid}{\bar{F}^\text{rigid}}
\newcommand*{\Trigid}{T^\text{rigid}}
\newcommand*{\rbar}{\bar{r}}

\newcommand*{\xiT}{\xi^\text{T}}

\newcommand*{\xiR}{\xi^\text{R}}
\newcommand*{\rcut}{r_\text{cut}}
\newcommand*{\Us}{U_\text{stretch}}
\newcommand*{\Ubend}{U_\text{bend}}
\newcommand*{\gammaves}{\xi_{\text{V}}}
\newcommand*{\noiseves}{\bar{\eta}^\text{ves}}
\newcommand*{\Uves}{U_\text{ves}}
\newcommand*{\Nves}{N_\text{ves}}
\newcommand*{\Nrods}{N_\text{rods}}

\newcommand*{\vcom}{\bar{v}_\text{com}}
\newcommand*{\fnet}{\bar{F}_\text{net}}
\newcommand*{\lv}{\lambda_\text{v}}
\newcommand*{\lf}{\lambda_\text{f}}

\newcommand*{\NrodsMax}{\Nrods^\text{max}}
\newcommand*{\BMMax}{\BM^\text{max}}

\begin{document}

\title{Design principles for transporting vesicles with enclosed active particles}
\begin{abstract}
We use coarse-grained molecular dynamics simulations to study the motility of a 2D vesicle containing self-propelled rods, as a function of the vesicle bending rigidity and the number density, length, and activity of the enclosed rods. Above a threshold value of the rod length, distinct dynamical regimes emerge, including a dramatic enhancement of vesicle motility characterized by a highly persistent random walk. These regimes are determined by clustering of the rods within the vesicle; the maximum motility state arises when there is one long-lived polar cluster. We develop a scaling theory that predicts the dynamical regimes as a function of control parameters, and shows that feedback between activity and passive membrane forces govern the rod organization. These findings yield design principles for building self-propelled superstructures using independent active agents under deformable confinement.
\end{abstract}
\author{Sarvesh Uplap}
\author{Michael F. Hagan}
\email{hagan@brandeis.edu}
\author{Aparna Baskaran}
\email{aparna@brandeis.edu}
\affiliation{Martin Fisher School of Physics, Brandeis University,\\
415 South Street, Waltham, Massachusetts 02453, USA}
\date{\today}
\maketitle

\section{Introduction}
Interacting systems of minimal motile agents occur on diverse length scales in biology, from the molecular motors and filaments in the cytoskeleton of the cell \cite{Klumpp2005, Furuta2012, Ross2008} to insect colonies \cite{gordon1999ants,costa2006other,detrain1999, buhl2006} and animal herds \cite{Ballerini2008, Silverberg2013, Garcimart2015, Cavagna2014}. The  emergent behaviors exhibited by these systems have been extensively studied within the paradigm of active matter \cite{RamaswamyReview2010,MarchettiRevModPhys2013,ReichhardtARCMPRev2017,Menzel2015,BechingerReview2016,MarchettiCurrOp2016,TailleurCatesRev2015,Vicsek2012}. In parallel, the same phenomena have inspired the field of collective robotics and swarm intelligence \cite{holland1999stigmergy,hamann2018swarm,tan2015handbook,kernbach2013handbook}, where it has been recognized that collective efforts of simple agents yield robust and  adaptable emergent behaviors that can be harnessed for specific tasks. Recent realizations of collective robotics have been used for programmable self-assembly \cite{Rubenstein2014,Gro2006,Yim2007,Goldstein2005}, mimicking foraging behavior of insects \cite{vaughan2008adaptive,parker2008multiple}, and sentry duties over unmapped and variable terrains \cite{Konolige2006}.

Physical interactions between motile agents will inevitably affect their collective behaviors. To successfully use minimally programmed agents to perform specific tasks, and to build small scale soft robotic systems, we should  leverage physical interactions as a design and control asset rather than a bottleneck that needs to be eliminated by programming or material design. Knowledge and models from statistical and soft matter physics can play important roles in this context. To that end, we consider a minimal example of collective behavior of motile agents ---  self-propelled (active) agents confined in an elastic vesicle. The specific function we consider is  transport of the vesicle, and we identify the role played by physical interactions in aiding or hindering this task.

Collections of confined active agents form transient clusters that push against their boundaries \cite{Wensink2008,Yang2014,Lee2013,Elgeti2009,Elgeti2013,caprini2018,Fily2015,Hiraoka2017,Levine2000,Peterson2021}. When confined to deformable boundaries, spontaneous fluctuations can result in the boundary developing large curvatures and spontaneous motion of the entire system \cite{Nikola2016,Paoluzzi2016,Tian_2017,Li2019,Wang2019}.
The benefit of such transient assemblies was demonstrated for a system of small simple robots confined by a flexible mobile enclosure \cite{Deblais2018}.
It was shown that aggregation of the robots led to transient directed motion in random directions and that the small robots can transport the enclosing frame around obstacles and through narrow openings, much like a collection of ants can manipulate large food items \cite{Feinerman2018, Reuveni2016}. However, to leverage this basic physics to build a robust collective robotic system, we must understand how spontaneous fluctuations can be preferentially biased and rectified so as to predominantly aid in the desired transport.

In this paper, we employ Langevin dynamics simulations to study the transport of a 2D elastic vesicle by self-organization of self-propelled rods enclosed within. The rods spontaneously form clusters on the vesicle boundary, and under certain conditions the rods form a single polar cluster which leads to highly efficient directed motion of the vesicle and its contents. However, under other conditions the rods form multiple clusters whose self-propulsion forces partially cancel, hindering motility. We find that the cluster organization is determined by a cooperative feedback between effective rod-rod attractions arising from their self-propulsion, the active forces of rods pushing on the vesicle boundary, and the passive reaction force from the deformed vesicle. Thus, the rod organization and consequent properties of the vesicle motion are determined by the number, aspect ratio, and self-propulsion velocity of the enclosed rods, as well as the deformability of the vesicle. We present simple scaling arguments that capture many of the observations from the simulations, and thereby identify  optimal design principles for maximizing the vesicle motility.

\section{Model}

We perform Langevin dynamics simulations in 2D of $\Nrods$  self-propelled, rigid rods enclosed by a passive elastic vesicle, which is represented as a semiflexible bead-spring ring polymer. Details of the microscopic model are as follows.

\textit{Self-propelled rods: }
Each rod consists of $n+1$ beads of diameter $\sigma$, with the centers of neighboring beads separated by a distance $b = 0.5\sigma$ so that the rod length is $\ell = nb + \sigma$.  Overlapping beads in this manner reduces surface roughness, thereby preventing interlocking of rods at high density \cite{Elgeti2013, Isele-Holder2015, Isele-Holder2016, Duman2018,Chelakkot2020,Peterson2021}.
The central bead is subjected to a constant self-propulsion force of magnitude $\fa$ along the rod axis, characterized by an orientation vector $\hat{\nu} = (\cos\theta, \sin\theta)$, where $\theta$ is the angle between the rod axis and the $x$ direction in the laboratory frame. Each rod bead interacts with all beads on other rods and all vesicle beads through a force-shifted WCA potential, $\ULJ = 4\epsilon \left[  \left(\frac{\sigma}{r}\right)^{12} - \left(\frac{\sigma}{r}\right)^6 \right] + \Delta V$, for interparticle distances $r<\rcut$ and zero otherwise ~\cite{Weeks1971, Andersen1971}. The shift in the potential is given by $\Delta V = -(r - \rcut)\frac{\partial \ULJ}{\partial r} |_{r = \rcut}$, the potential is cutoff at the Lennard-Jones minimum distance $\rcut =2^{1/6} \sigma$, and the repulsion strength is set to $\epsilon=\kT$. 

The mass of the central bead, which is set to $m=1$, sets the total rod mass. The equations of motion are integrated for the central bead of each rod only (the rods move as rigid bodies), with the forces $\Frigid$ and torques $\Trigid$ from the constituent particles transferred to the central bead. The equations of motion for the $i^{th}$ rod are:
\begin{equation}
  \xiT \frac{d\rbar_i}{dt} = \fa\hat{\nu} + \Frigid_i + \noiseT_i(t)
\end{equation}

\begin{equation}
   \xiR \frac{d\theta_i}{dt} =  \Trigid_i + \noiseR_i(t)
\end{equation}
\begin{align}
 \Frigid_i = & -\sum_{j\ne i}^{\Nrods} \sum_{m,m'=1}^n \nabla_{\rbar_i}  \ULJ\left(|\rbar_{i,m}-\rbar_{j,m'}|\right) \nonumber \\
  & -\sum_{k=1}^{\Nves} \sum_{m=1}^n \nabla_{\rbar_i} \ULJ\left(|\rbar_{i,m}-\rbar_{k}|\right)
  \label{eq:Frigid}
\end{align}
where   $\xiT = 10(n+1)\sqrt{\kT}\sigma$, $\xiR = 1000\sqrt{\kT}/\sigma$ are the translational and rotational drag coefficients for each rod; $\noiseT_i$, $\noiseR_i$ are uniform random forces modelled as Gaussian white noise with moments $\langle \noiseT_i(t) \rangle = 0$, $\langle \noiseR_i(t) \rangle = 0$ and $\langle \noiseT_i(t) \cdot \noiseT_{j}(t') \rangle = 4\kT \xiT \delta_{i,j}\delta(t-t')$; and $\langle \noiseR_i(t) \cdot \noiseR_{j}(t') \rangle = 2\kT \xiR \delta_{i,j}\delta(t-t')$. In Eq.~\eqref{eq:Frigid}, $\Nrods$ is the number of rods in the system, $\rbar_{i,m}$ is the position of bead $m$ on rod $i$, and $\rbar_{k}$ is the position of vesicle bead $k$.  The torque $\Trigid_i$ is computed from the forces on each bead in rod $i$.

\textit{Elastic vesicle}:
We model the vesicle as a passive semiflexible bead-spring ring polymer. As for the filaments, the bead diameter is $\sigma$ and the equilibrium distance between neighboring beads is $b = 0.5\sigma$. The total interaction potential for the vesicle is given by
\begin{align}
\Uves = & \sum_{i=1}^{\Nves-1} \Us(r_{i,i+1}) + \nonumber \\
& \sum_{i=2}^{\Nves-1}  \Ubend(\phi_{i-1,i,i+1}) + \sum_{\langle ij \rangle} \ULJ(r_{ij})
\label{eq:Uves}
\end{align}
where $r_{i,j}\equiv|\rbar_i-\rbar_{j}|$; stretching penalties are enforced by harmonic bonds between neighboring beads, $\Us(r)=\sm(r-b)^2$ with $\sm$ the stretching modulus; the bending energy is given by $\Ubend(\phi)=\BM (\pi - \phi)^2$, with $\phi$ the angle made by three consecutive beads (or two consecutive bonds) and $\BM$ as the bending modulus. The non-bonded interaction is given by the WCA potential 
 and is summed over all vesicle-rod pairs of beads. 

The equation of motion for the $i^{th}$ vesicle bead is:
\begin{equation}
    \frac{d^2\rbar_i}{dt^2} =  -\gammaves \frac{d{\rbar}_i}{dt} - \nabla_{{\rbar}_i} \Uves + \noiseves_i(t)
\end{equation}
where $\gammaves = \sqrt{\kT}\sigma$ is the translational drag coefficient and $\noiseves_i$ is Gaussian white noise representing thermal motions. The moments of the thermal noise are $\langle \noiseves_i(t) \rangle = 0$ and $\langle \noiseves_i(t) \cdot \noiseves_j(t')\rangle = 4\kT \gammaves \delta_{i,j}\delta(t-t')$. The mass of each vesicle bead is set to $m=1$. For all results reported in this work the vesicle comprises $\Nves=1005$ beads so that the equilibrium vesicle radius is $\Rv=80\sigma$.

\begin{figure}[hbt]
\centering
\includegraphics[width=0.475\textwidth]{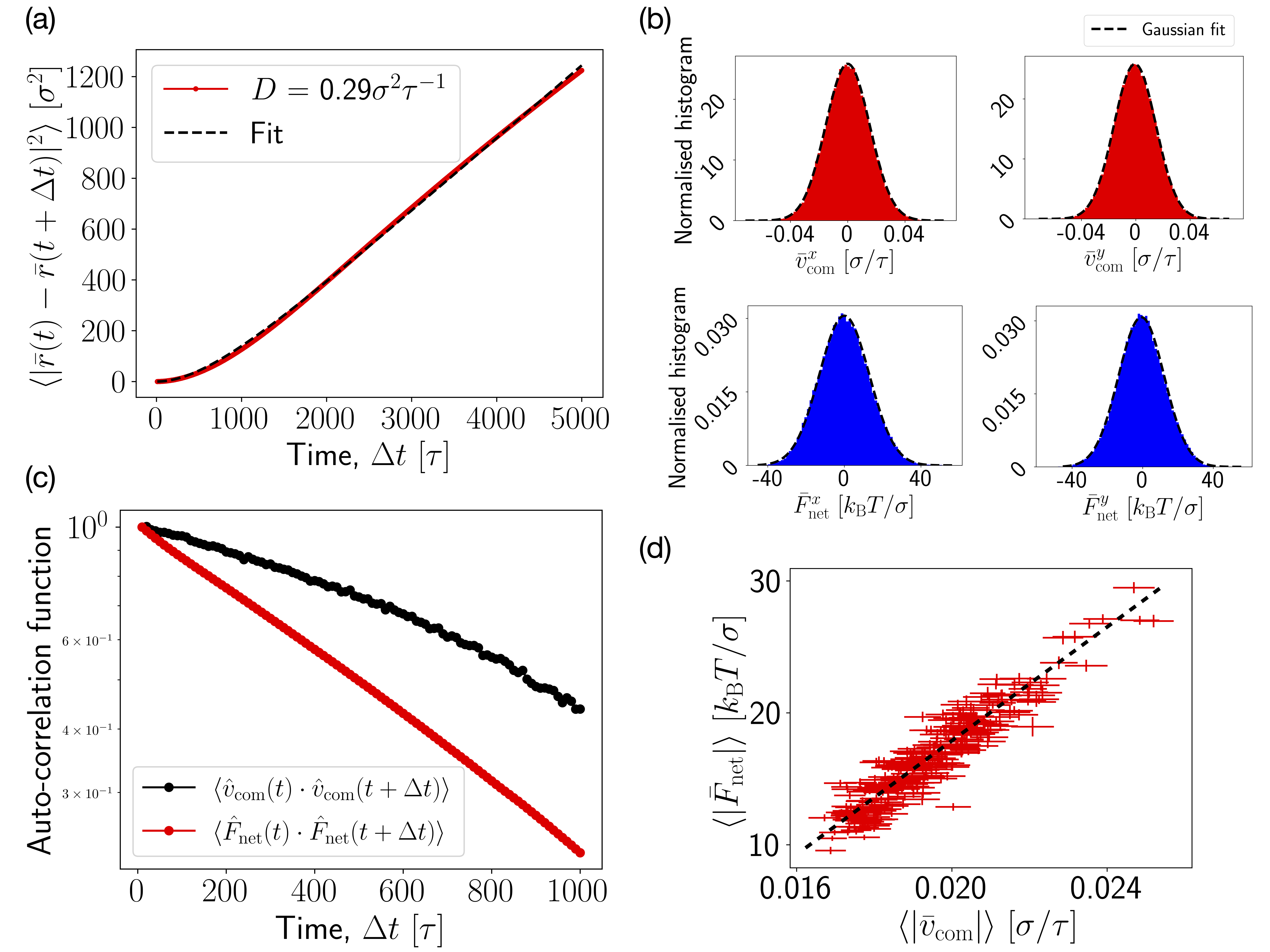}
\caption{Statistical properties of the center-of-mass motion of a vesicle with bending modulus $\BM=3000\kT$ containing $\Nrods=30$ rods of aspect ratio $\ell/\sigma=3$ with active force $\fa=3\kT/\sigma$. (a) The mean square displacement of the center of mass, fit to the functional form of an active Brownian particle to obtain the translational diffusion coefficient $D$. (b) The histogram of $x$ and $y$ components of the net force $\fnet$ and center-of-mass velocity $\vcom$. The distributions are Gaussian with zero mean. (c) Autocorrelation functions for the net force and velocity, $\left\langle \fnet\left( t\right) \cdot \fnet\left( t+\Delta t\right)\right\rangle $ and $\left\langle \vcom(t)\cdot \vcom(t+\Delta t)\right\rangle$. The  correlation times $\lv$ and $\lf$ are extracted by fitting these measurements to exponentials. (d) Plot of the mean active force $\langle|\fnet|\rangle$ against the mean velocity,  $\langle |\vcom|\rangle$, showing they are highly correlated.
}
\label{fig:1}
\end{figure}

\section{Results}
\subsection{Phenomenology of vesicle motion}
Our goal is to characterize the motility of the vesicle. We begin by focusing on its center of mass,  $\bar{r}(t)$. We find that the dynamics of $\bar{r}(t)$ is well described by a persistent random walk with $\langle |\bar{r}(t)-
\bar{r}(t+\Delta t)|^{2}\rangle=D[\Delta t-\lambda +\lambda
e^{-\Delta t/\lambda }]$, where $\lambda$ is the persistence time and $D$ is the diffusion coefficient describing the vesicle motions at times $t\gg\lambda$ (Fig. \ref{fig:1}a).  Further, we measure $\fnet$, the net active force on the vesicle, exerted by the enclosed rods that are in contact with it and  $\vcom$, the center of mass velocity. We find that both quantities are  distributed normally (Fig. \ref{fig:1}b) and exponentially correlated in time (Fig.  \ref{fig:1}c). Finally, the dynamics of the vesicle center of mass is reliably overdamped (Fig.  \ref{fig:1}d). Thus, the composite system of the vesicle and its enclosed rods effectively behave as an Active Brownian Particle. This is the conceptual framework we will use to characterize the vesicle motility as we change the physical properties of its building blocks.

\subsection{Dependence of motility on vesicle stiffness and active rod length}
\label{sec:motility}
To highlight the key findings of this study, let us consider the physics we know  from prior studies on self-propelled particles at curved walls \cite{Fily2015,Fily2014,Tian2021, Nikola2016, Smallenburg2015, Li2019, Paoluzzi2016,Tian_2017,Li2019,Wang2019}. Self-propelled particles cluster at walls \cite{Wensink2008,Elgeti2009,Elgeti2013,Yang2014,Elgeti2015}, and  tend to accumulate in regions of high curvature \cite{Fily2014, Mallory2014, Fily2015, Vutukuri2019}. Since our vesicle is deformable in the present work, we expect curvature to increase in regions of accumulation and thereby lead to recruitment of more particles into the cluster\cite{Nikola2016,Paoluzzi2016,Tian_2017,Li2019,Wang2019,Deblais2018,Vutukuri2019}. This feedback mechanism should be greater for vesicles with lower bending rigidity, since the curvature amplification will be larger for a given active force magnitude and local particle density. Thus, we might naively expect softer vesicles to have higher motility compared to stiff ones, since they provide a pathway to the formation of a few larger clusters rather than a more uniform distribution of particles along the boundary.

\begin{figure}[hbt]
\centering
\includegraphics[width=0.475\textwidth]{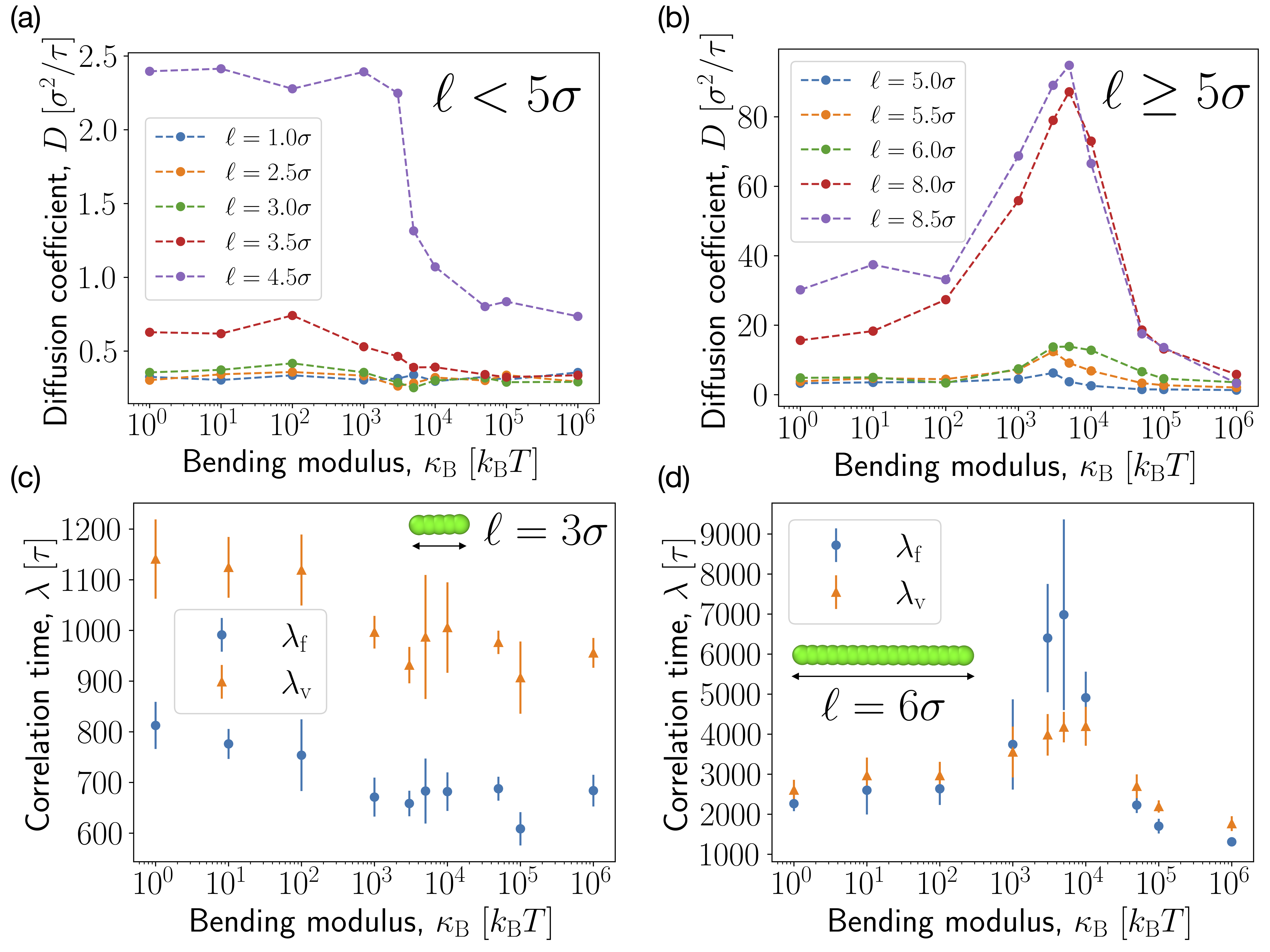}
\caption{(a,b) Dependence of the vesicle diffusion constant on the vesicle bending modulus $\BM$ for indicated values of the rod aspect ratio $\ell/\sigma$, for aspect ratios that are (a) below the threshold for long-lived clusters ($\ell/\sigma < 5$) and (b) above the threshold ($\ell/\sigma \ge 5$).  (c,d) The correlation times of the net force and the center-of-mass velocity for (c) below threshold aspect ratio ($\ell/\sigma=3$), showing that both are independent of  vesicle stiffness, and (d) above threshold ($\ell/\sigma=6$) showing enhanced correlation at the optimal stiffness.}
\label{fig:2}
\end{figure}

Our first observation is that this simple argument is not the full story. Fig.~\ref{fig:2} characterizes the vesicle motility for fixed $\Nrods=30$ rods and activity $\fa=3\kT/\sigma$  as a function of the rod aspect ratio $\ell/\sigma$ and vesicle stiffness  $\BM$. We identify three distinct behaviors: (i) for $\ell/\sigma \leq 3$, the motility is independent of vesicle bending stiffness, (ii) for $3<\ell/\sigma\leq 5$, the motility increases as the vesicle softens, and (iii) for longer rods a strong non-monotonic dependence on stiffness emerges. 

We can understand these changes in phenomenology as a function of aspect ratio by considering the limiting case of $\ell/\sigma=1$, i.e., spheres, in an infinitely stiff, i.e., rigid, circular vesicle. Supposing that the system was initialized with isotropically distributed self-propulsion forces, as all the simulations in this study are, we expect a uniform distribution of particle orientations on the vesicle boundary. Therefore, the net force on the center of mass is small and its fluctuations are correlated on the same time scale as the bare correlations of isolated active particles.  Hence, the enclosed active particles will not enhance the vesicle motility to any significant degree. When we move away from the rigid limit, deformations of the vesicle do induce larger clusters and significant density variation along the boundary. But, as $\ell/\sigma\to1$, neither the boundary nor the inter-particle interactions  can exert significant torques to realign the active forces. Thus, the net force on the vesicle remains small and enhancement of motility negligible. Given this picture, we can understand the low aspect ratio results  in Fig.~\ref{fig:2}a as the generalization of this asymptotic case of spherical particles.

Now let us switch focus to the stiffness dependence of motility for longer rods. Again, it is useful to first build a heuristic based on known phenomenology \cite{Br2020}. We first consider long rods accumulated at a rigid circular boundary. Suppose the rods do not interact with each other, i.e., in the dilute limit, they would slide along the boundary until they become tangential to it \cite{Fily2014, Fily2015, Elgeti2013, BechingerReview2016, Wensink2008}. But, when the rods collide at the boundary, the inter-particle interactions  will lead the rods to align such that they form polar clusters that are normal to the boundary \cite{Peterson2021, Wensink2008, BechingerReview2016}. These polar clusters push on the boundary, thus inducing and amplifying local curvature fluctuations. In turn, the rods' forward motion couples to the vesicle curvature to result in an effective attraction between neighboring rods. For softer vesicles, the induced curvature and corresponding effective rod-rod attraction increases, thus resulting in larger and longer-lived polar clusters. For moderate aspect-ratio rods, this effect leads to an increase in vesicle motility with decreasing vesicle rigidity (most easily seen in Fig.~\ref{fig:2}a for $\ell/\sigma=4.5$). However, for longer rods, thisphenomenology is overwhelmed by the emergence of an optimal stiffness, around which there is a huge enhancement of motility above the basic trend (see Fig.~\ref{fig:2} (b,d)).

\begin{figure}[hbt]
\centering
\includegraphics[width=0.475\textwidth]{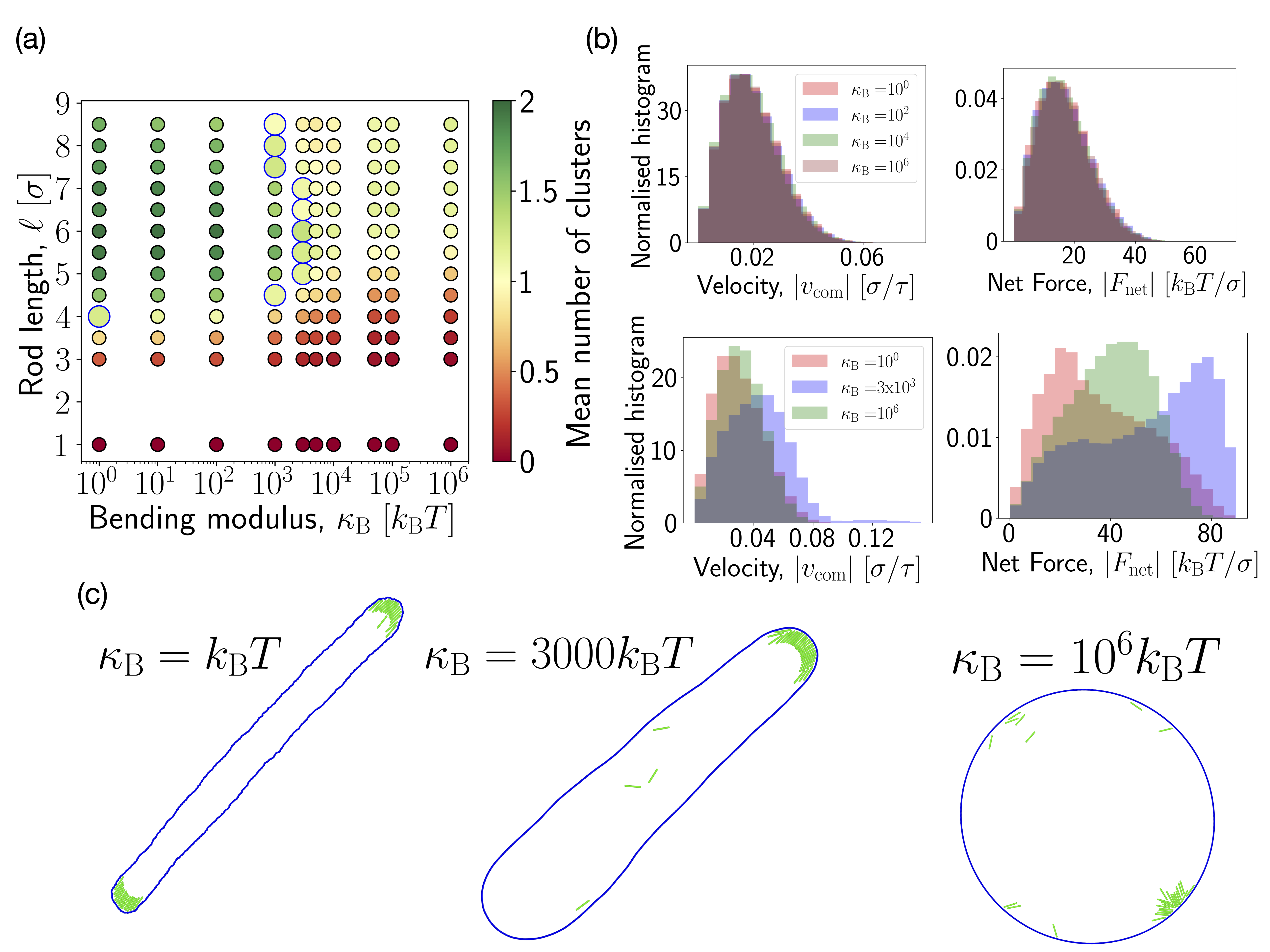}
\caption{(a) Mean number of clusters as a function of rod aspect ratio and vesicle bending modulus for active force $\fa=3\kT/\sigma$. A single long-lived cluster emerges at the optimal value of stiffness,  $\BM \approx 3000\kT$. The optimal stiffness for each aspect ratio is indicated by an enlarged marker. (b) Histograms of $\fnet$ and $\vcom$ for (top) $\ell/\sigma =3$ and (bottom) $\ell/\sigma=6$, showing that the net force shifts to large values at the optimal stiffness for long rods.  (c) Snapshots illustrating the rod organization for $\ell/\sigma = 10$ for  (left) low, (middle) optimal, and (right) high bending modulus.}
\label{fig:3}
\end{figure}

To elucidate the mechanism underlying the dramatic motility enhancement for optimal vesicle stiffness, we now examine the organization of active rods within the vesicle (Fig. \ref{fig:3}). We find that the enhanced motility states are characterized by the presence of a single (Fig. \ref{fig:3}a) long-lived (SI Fig.~5) polar cluster of rods. Hence, essentially all encapsulated rods point in the same direction, leading to a large net active force on the vesicle $\fnet \thicksim \fa \Nrods$ and correspondingly a large persistence length for the vesicle's motion. 
For vesicles that are softer than the optimal stiffness, the rods organize into two or more polar clusters. Through the combination of the clusters’ polar motion and the consequent local curvature of the membrane, there is an effective repulsion between clusters driving them into a steady-state configuration in which their active forces tend to cancel (see Fig. \ref{fig:3}c,  Fig. \ref{fig:5}a, and SI section I). Consequently, the vesicle exhibits a much smaller persistence length than for the 1-cluster state.
For stiffer-than-optimal vesicles, the rods can only weakly deform the vesicle boundary. Since local curvature of the vesicle is essential to stabilize the polar clusters (see below), only transient clusters form. Therefore, the direction of $\fnet$ fluctuates rapidly and the vesicle motion is characterized by a small persistence length. These observations suggest that both the aligning interactions among the rods due to their self-propulsion and vesicle curvature are crucial to the emergence of the enhanced motility states. As evidence of the importance of direct rod-rod alignment interactions, we observe large net active forces on the vesicle only for large aspect-ratio rods (Fig.~\ref{fig:3}b).

\subsection{Assembly principles for the single cluster state}
\label{sec:assemblyPrinciples}
Now, we seek to quantify the emergence of a single cluster in terms of the physical parameters of the building blocks of our motile vesicle. As we have shown in previous work \cite{Peterson2021}, long-lived clusters can be fruitfully described as  self-limited structures whose size can be accurately captured using the concepts of self-assembly.  We lay out below a physical description of the model that predicts the number and sizes of clusters as a function of parameters, and we use these estimates to predict the optimal stiffness at which the huge enhancement in motility occurs.  The mathematical details are given in SI Section I.  

\begin{figure}[hbt]
\centering
\includegraphics[width=0.475\textwidth]{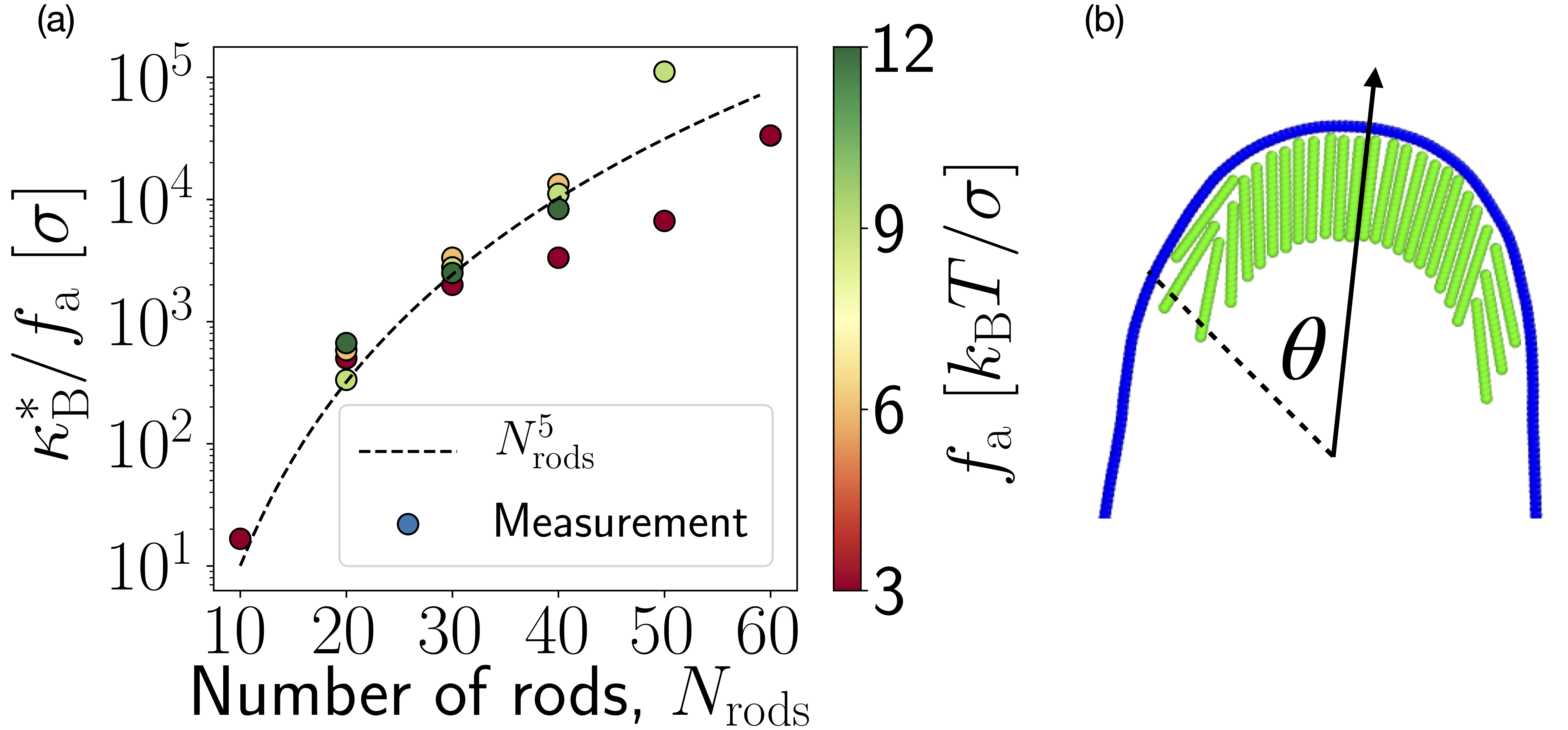}
\caption{(a) Comparison of the theoretical scaling of the optimal bending modulus with the number of enclosed rods and active force, $\BM^*/\fa\sim \Nrods^{5}$, with measurements from the simulations. (b) Schematic showing a visual representation of the angular size $\theta$ of a cluster.  }
\label{fig:4}
\end{figure}

We know from previous work that self-propulsion together with the presence of the wall drives rods to align with each other, perpendicular to the wall, and to maximize the overlap along their length with their neighbors. That is, the rods tend to form smectic layers at the wall \cite{Br2020}. We can capture this phenomenology through an effective `energy' of the form $U_{\text{active}}+U_{\text{interfacial}}$, where $U_{\text{active}}=-C\sigma \fa n_{\text{r}}$, an attractive interaction that scales with the activity and number of rods in the cluster $n_{\text{r}}$, and $U_{\text{interfacial}}=-2\gamma$ is a surface tension that accounts for the absence of neighbors at either edge of a cluster. The surface tension encodes the tendency for clusters to grow. While this captures the phenomenology at a flat wall, the curvature of the vesicle `opposes' the growth of the cluster by forcing rods within a cluster to shear relative to their neighbors, thus preventing the perfect overlap that flat walls would allow.
We can capture this effect through an energy $U_{\text{shear}}=-2k_{\text{shear}}\log(\cos\theta)$, where $\theta$ is the angular extent of the cluster (see Fig.~\ref{fig:4}b) and the constant $k_{\text{shear}}$ also scales with activity $\fa$, consistent with the attractive interactions that lead to the clustering. 

For a given cluster size, the angular extent of a cluster depends on the local radius of curvature of the vesicle, which is determined by a balance of active force of the rods and the elastic response of the vesicle. We describe this phenomenology through an energy $U_{\text{ves}}=U_{\text{bend}}+U_{\text{f}}$, where $U_{\text{bend}}$ is the standard Helfrich free energy \cite{Helfrich1973,Deserno2015,Saric2013} and $U_{\text{f}}=-\fa n_{\text{r}}l$ is the work done by the active force (see SI Section I). Putting all this together, the number and size of clusters, and correspondingly the steady-state geometry of the vesicle, can be described by finding the minimum of a free energy of the form $U(\BM,\fa,\Nrods)=U_{\text{rods}}+U_{\text{ves}}$, where $U_{\text{rods}}=U_{\text{active}}+U_{\text{interfacial}}+U_{\text{shear}}$, described above.

In particular, the analysis shows that, starting from the floppy vesicle limit $\BM\to 0$, the number of stable clusters decreases with increasing bending modulus.  By calculating when the system transitions from a two-cluster state to a single cluster state, corresponding to the highly motile states observed in the simulations, we can estimate that the optimal stiffness scales as $\BM^*/\fa\sim \Nrods^{5}$ (see Eq. 20 in section ID of the SI). The predicted scaling with both activity and number of enclosed rods and shows good agreement with our simulation results (Fig.~\ref{fig:4}a), although the accessible range of $\Nrods$ within the vesicle size that we focus on ($\Rv=80\sigma$) is insufficient to rigorously test the scaling exponent. 

The analysis predicts that this trend continues until a threshold value of the bending modulus $\BMMax\sim\Rv^2\fa$, above which the active force is insufficient to deform the membrane. Because membrane deformation is essential to stabilize long-lived polar cluster states, we expect only transient clusters for stiffer vesicles. Furthermore, equating this limit with the transition value for the single cluster state, i.e. $\BM^*=\BMMax$, identifies a maximum number of enclosed rods $\NrodsMax\sim \left(\Rv \ell\right)^{2/5}$ above which the system will transition directly from multiple-cluster states to transient clusters with increasing bending modulus. Thus, for $\Nrods > \NrodsMax$ the system will not exhibit the single-cluster high-motility state for any parameter values. For our vesicle size $\Rv=80\sigma$  and the range of aspect-ratios that lead to polar clusters ($\ell/\sigma\ge6$) this estimates $\NrodsMax$ is order 10 to within a scaling constant, which is consistent with the simulation results that we do not observe the single-cluster state for $\Nrods\gtrsim 70$.

Since the number of clusters is closely linked to the steady-state geometry of the vesicle configuration (see  Fig.~\ref{fig:5}), further analysis of this theoretical model can be leveraged to design shapes of 2D active vesicles (similar to \cite{Peterson2021, Vutukuri2019}). We defer this to future work, in order to keep the focus of this article on motility of the vesicle.

\subsection{Dependence of rod organization and vesicle motility on number of enclosed rods}
\label{sec:Nrod}
\begin{figure}[hbt]
\centering
\includegraphics[width=0.475\textwidth]{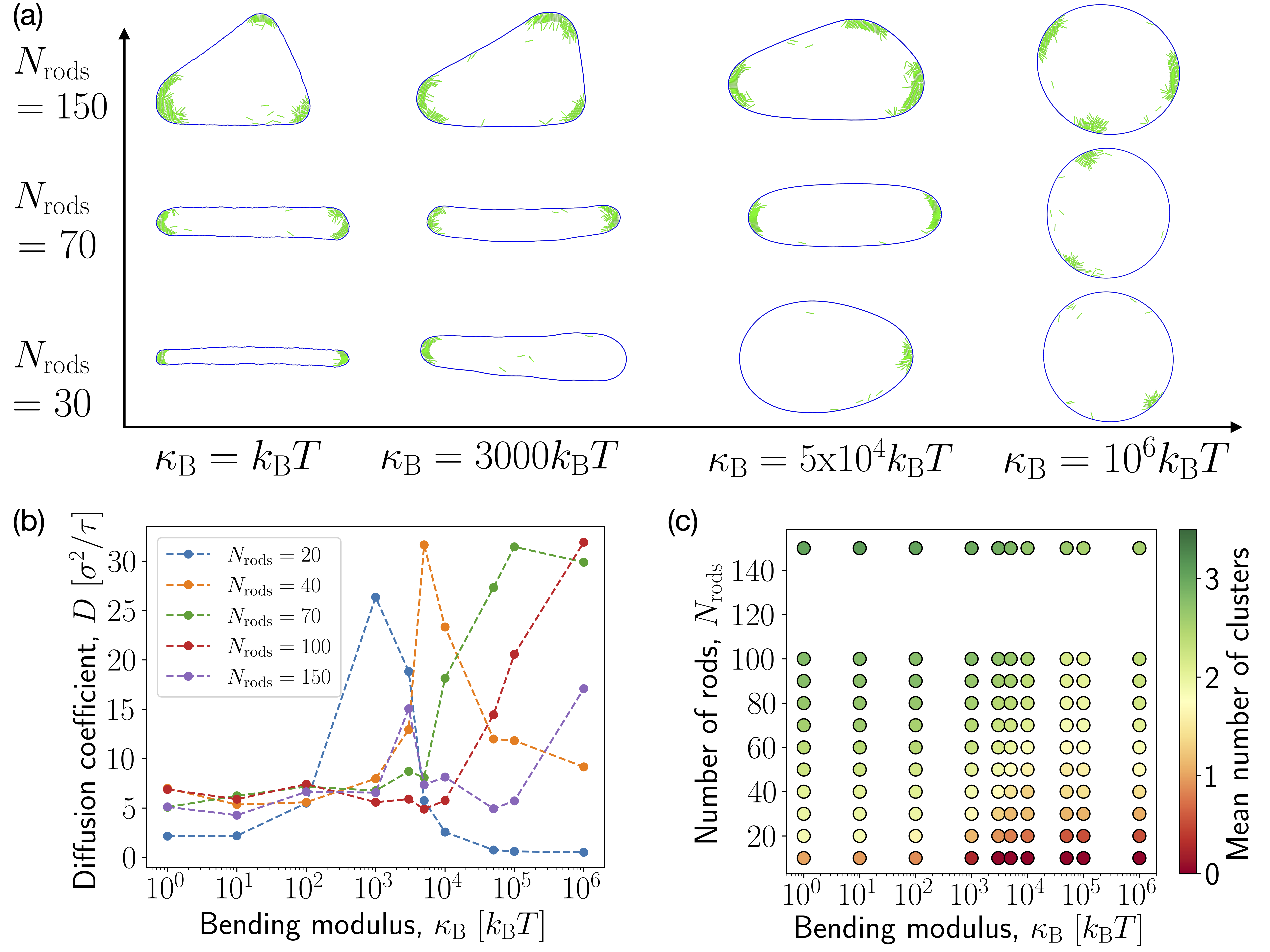}
\caption{ (a) Snapshots of states that typify the dominant cluster configurations as a function of number of enclosed rods $\Nrods$ and vesicle bending modulus. (b) The vesicle diffusion coefficient as a function of bending modulus for indicated $\Nrods$. (c) The mean number of clusters within a vesicle as a function of bending modulus and $\Nrods$, showing that the single-cluster state is not observed for $\Nrods\gtrsim 70$ enclosed rods. Results are shown for $\fa=3\kT/\sigma$ and $\ell/\sigma=10$.}
\label{fig:5}
\end{figure}

The scaling relationship above provides a quantitative design principle for building highly motile vesicles. But one important limitation it identifies is that motility enhancement through the formation of the single cluster state only applies in the dilute limit.  This corresponds to $\Nrods<\NrodsMax \approx 70$ for the vesicle size that we focus on $\Rv=80\sigma$. Above this limit, there are too many rods to form a state in which there is a single polar-aligned cluster. 
Consequently, the simple picture of three distinct dynamical regimes for large aspect-ratio rods ($\ell>5\sigma$) breaks down (Fig.~\ref{fig:5}). 

Fig. \ref{fig:5}a is a visual representation of the structures formed by the enclosed rods as the number of rods increases. In the intermediate regime, ($70 \lesssim \Nrods \lesssim 100$ for $\Rv=80\sigma$), the dominant steady-state configurations gradually shift from 2-cluster to 3-cluster states with increasing $\Nrods$ in floppy vesicles, 1-cluster to 2-cluster states for intermediate stiffness, and one or more transient clusters at high vesicle stiffness. However, even though self-limited clusters continue to form in softer vesicles,  the clusters lack perfect polar alignment and tend to become multilayered. 
We speculate that this phenomenology reflects the onset of motility induced phase separation (MIPS) related behaviors at these densities \cite{Br2020, Peruani2011, Weitz2015, vanderLinden2019}, which unevenly and transiently enhance clustering.  At higher densities ($\Nrods>100$ for $\Rv=80\sigma$ and $\fa=3\kT/\sigma$),  the MIPS dominates the organization of the self-propelled rods at the boundary. In this regime, the rods tend to form \emph{aggregates} comprising multiple sub-clusters with different orientations. For example, in floppy vesicles, the dominant rod organization corresponds to three aggregates sitting at three vertices of a triangular configuration. As noted above, the combination of rod-propulsion and membrane-curvature-mediated interactions tends to drive clusters away from each other, stabilizing the triangular arrangement. However, because the rods within a given aggregate are not all aligned, aggregates are transient, tending to break apart and merge with other aggregates. With increasing $\BM$ the mean number of clusters decreases, but the clusters remain transient and thus do not lead to the highly motile vesicle dynamics characterized by a long persistence length. Furthermore, even in the 1-cluster states at these higher rod densities, there are additional rods forming transient clusters elsewhere on the boundary, which have orientations for which the propulsion forces partly cancel those of the rods in the cluster.

Fig. \ref{fig:5}b characterizes the motility of the vesicle for different numbers of enclosed rods. For the intermediate densities, the diffusion coefficient as a function of vesicle stiffness shows similar trends as seen in Fig.~\ref{fig:2}b, with motility increasing with $\BM$ for floppy vesicles and the value of optimal stiffness $\BM^*$ shifting toward larger values with increasing number of rods. 
This is consistent with the fact that the number of clusters decreases with stiffness, thus increasing the average magnitude of the net active force. However, the motility enhancement is significantly smaller than for $\Nrods<\NrodsMax$,  due to the imperfect alignment of the clusters.
 At higher densities ($\Nrods>100$), the motility becomes largely independent of stiffness. This reflects the fact that, when the MIPS phenomenology dominates, the clusters lack polar order and are transient. Thus, the net active force lacks the large magnitude and long-lived orientation that drive directed vesicle motion.

\section{Summary and Outlook}
This work indentifies design principles for constructing elastic vesicles with programmable classes of motilities, which range from Brownian to highly persistent, by encapsulating self-propelled rods. In particular, for a given rod aspect ratio and activity level, our simple theoretical framework predicts the vesicle bending modulus and number of encapsulated agents that maximize the vesicle motility.  This framework builds on the well-established phenomenology of self-propelled particles under confinement, and is consistent with observations from our particle-based simulations. The theoretical framework can be applied to other functional goals, such as such as prescribing the shape of an elastic vesicle by enclosing active particles within it.

There is now a robust understanding of active matter phenomenology for many model systems. However, leveraging this fundamental knowledge to design architectures with specific functional goals has been relatively limited. We hope that this investigation, on the simple functional goal of emergent motility, can serve as a template for investigations into other, potentially more sophisticated functions. A direct extension of this work would be to leverage physical interactions or spatiotemporal activity to render an elastic vesicle steerable. For example, one could consider vesicles of nonuniform stiffness where cluster formation occurs preferentially at certain sites, or cycling activity as a function of time. Such investigations may lead to a better understanding of how to extract function from systems of interacting motile agents.

\begin{acknowledgements}
This work was supported by the National Science Foundation (NSF) DMR-1855914 (SU and MFH), DMR-2202353 (AB) and the Brandeis Center for Bioinspired Soft Materials, an NSF MRSEC DMR-2011846 (SU, AB, and MFH). Computing resources were provided by the NSF XSEDE/ACCESS allocation MCB090163 (Stampede and Expanse) and the Brandeis HPCC which is partially supported by NSF DMR-MRSEC 2011846 and OAC-1920147. 
\end{acknowledgements}

\bibliography{rsc}

\begin{thebibliography}{67}%
\makeatletter
\providecommand \@ifxundefined [1]{%
 \@ifx{#1\undefined}
}%
\providecommand \@ifnum [1]{%
 \ifnum #1\expandafter \@firstoftwo
 \else \expandafter \@secondoftwo
 \fi
}%
\providecommand \@ifx [1]{%
 \ifx #1\expandafter \@firstoftwo
 \else \expandafter \@secondoftwo
 \fi
}%
\providecommand \natexlab [1]{#1}%
\providecommand \enquote  [1]{``#1''}%
\providecommand \bibnamefont  [1]{#1}%
\providecommand \bibfnamefont [1]{#1}%
\providecommand \citenamefont [1]{#1}%
\providecommand \href@noop [0]{\@secondoftwo}%
\providecommand \href [0]{\begingroup \@sanitize@url \@href}%
\providecommand \@href[1]{\@@startlink{#1}\@@href}%
\providecommand \@@href[1]{\endgroup#1\@@endlink}%
\providecommand \@sanitize@url [0]{\catcode `\\12\catcode `\$12\catcode
  `\&12\catcode `\#12\catcode `\^12\catcode `\_12\catcode `\%12\relax}%
\providecommand \@@startlink[1]{}%
\providecommand \@@endlink[0]{}%
\providecommand \url  [0]{\begingroup\@sanitize@url \@url }%
\providecommand \@url [1]{\endgroup\@href {#1}{\urlprefix }}%
\providecommand \urlprefix  [0]{URL }%
\providecommand \Eprint [0]{\href }%
\providecommand \doibase [0]{https://doi.org/}%
\providecommand \selectlanguage [0]{\@gobble}%
\providecommand \bibinfo  [0]{\@secondoftwo}%
\providecommand \bibfield  [0]{\@secondoftwo}%
\providecommand \translation [1]{[#1]}%
\providecommand \BibitemOpen [0]{}%
\providecommand \bibitemStop [0]{}%
\providecommand \bibitemNoStop [0]{.\EOS\space}%
\providecommand \EOS [0]{\spacefactor3000\relax}%
\providecommand \BibitemShut  [1]{\csname bibitem#1\endcsname}%
\let\auto@bib@innerbib\@empty
\bibitem [{\citenamefont {{S. Klumpp, R. Lipowsky}}(2005)}]{Klumpp2005}%
  \BibitemOpen
  \bibfield  {author} {\bibinfo {author} {\bibnamefont {{S. Klumpp, R.
  Lipowsky}}},\ }\bibfield  {title} {\bibinfo {title} {Cooperative cargo
  transport by several molecular motors},\ }\href
  {https://doi.org/10.1073/pnas.0507363102} {\bibfield  {journal} {\bibinfo
  {journal} {Proceedings of the National Academy of Sciences}\ }\textbf
  {\bibinfo {volume} {102}},\ \bibinfo {pages} {17284} (\bibinfo {year}
  {2005})}\BibitemShut {NoStop}%
\bibitem [{\citenamefont {{K. Furuta, A. Furuta, Y. Y. Toyoshima, M. Amino, K.
  Oiwa, H. Kojima}}(2012)}]{Furuta2012}%
  \BibitemOpen
  \bibfield  {author} {\bibinfo {author} {\bibnamefont {{K. Furuta, A. Furuta,
  Y. Y. Toyoshima, M. Amino, K. Oiwa, H. Kojima}}},\ }\bibfield  {title}
  {\bibinfo {title} {Measuring collective transport by defined numbers of
  processive and nonprocessive kinesin motors},\ }\href
  {https://doi.org/10.1073/pnas.1201390110} {\bibfield  {journal} {\bibinfo
  {journal} {Proceedings of the National Academy of Sciences}\ }\textbf
  {\bibinfo {volume} {110}},\ \bibinfo {pages} {501} (\bibinfo {year}
  {2012})}\BibitemShut {NoStop}%
\bibitem [{\citenamefont {{J. L. Ross, M. Y. Ali, D. M.
  Warshaw}}(2008)}]{Ross2008}%
  \BibitemOpen
  \bibfield  {author} {\bibinfo {author} {\bibnamefont {{J. L. Ross, M. Y. Ali,
  D. M. Warshaw}}},\ }\bibfield  {title} {\bibinfo {title} {Cargo transport:
  molecular motors navigate a complex cytoskeleton},\ }\href
  {https://doi.org/10.1016/j.ceb.2007.11.006} {\bibfield  {journal} {\bibinfo
  {journal} {Current Opinion in Cell Biology}\ }\textbf {\bibinfo {volume}
  {20}},\ \bibinfo {pages} {41} (\bibinfo {year} {2008})}\BibitemShut {NoStop}%
\bibitem [{\citenamefont {{Gordon, M. Deborah}}(1999)}]{gordon1999ants}%
  \BibitemOpen
  \bibfield  {author} {\bibinfo {author} {\bibnamefont {{Gordon, M.
  Deborah}}},\ }\href@noop {} {\emph {\bibinfo {title} {Ants at work: how an
  insect society is organized}}}\ (\bibinfo  {publisher} {Simon and Schuster},\
  \bibinfo {year} {1999})\BibitemShut {NoStop}%
\bibitem [{\citenamefont {{Costa, T. James}}(2006)}]{costa2006other}%
  \BibitemOpen
  \bibfield  {author} {\bibinfo {author} {\bibnamefont {{Costa, T. James}}},\
  }\href@noop {} {\emph {\bibinfo {title} {The other insect societies}}}\
  (\bibinfo  {publisher} {Harvard University Press},\ \bibinfo {year}
  {2006})\BibitemShut {NoStop}%
\bibitem [{\citenamefont {{C. Detrain, J. L. Deneubourg, J. M. Pasteels, J. M.
  Pasteels}}(1999)}]{detrain1999}%
  \BibitemOpen
  \bibfield  {author} {\bibinfo {author} {\bibnamefont {{C. Detrain, J. L.
  Deneubourg, J. M. Pasteels, J. M. Pasteels}}},\ }\href@noop {} {\emph
  {\bibinfo {title} {Information processing in social insects}}}\ (\bibinfo
  {publisher} {Springer Science Business Media},\ \bibinfo {year}
  {1999})\BibitemShut {NoStop}%
\bibitem [{\citenamefont {{J. Buhl, D. J. T. Sumpter, I. D. Couzin, J. J. Hale,
  E. Despland, E. R. Miller, S. J. Simpson}}(2006)}]{buhl2006}%
  \BibitemOpen
  \bibfield  {author} {\bibinfo {author} {\bibnamefont {{J. Buhl, D. J. T.
  Sumpter, I. D. Couzin, J. J. Hale, E. Despland, E. R. Miller, S. J.
  Simpson}}},\ }\bibfield  {title} {\bibinfo {title} {From disorder to order in
  marching locusts},\ }\href {https://doi.org/10.1126/science.1125142}
  {\bibfield  {journal} {\bibinfo  {journal} {Science}\ }\textbf {\bibinfo
  {volume} {312}},\ \bibinfo {pages} {1402} (\bibinfo {year}
  {2006})}\BibitemShut {NoStop}%
\bibitem [{\citenamefont {{M. Ballerini, N. Cabibbo, R. Candelier, A. Cavagna,
  E. Cisbani, I. Giardina, V. Lecomte, A. Orlandi, G. Parisi, A. Procaccini, M.
  Viale, V. Zdravkovic}}(2008)}]{Ballerini2008}%
  \BibitemOpen
  \bibfield  {author} {\bibinfo {author} {\bibnamefont {{M. Ballerini, N.
  Cabibbo, R. Candelier, A. Cavagna, E. Cisbani, I. Giardina, V. Lecomte, A.
  Orlandi, G. Parisi, A. Procaccini, M. Viale, V. Zdravkovic}}},\ }\bibfield
  {title} {\bibinfo {title} {Interaction ruling animal collective behavior
  depends on topological rather than metric distance: Evidence from a field
  study},\ }\href {https://doi.org/10.1073/pnas.0711437105} {\bibfield
  {journal} {\bibinfo  {journal} {Proceedings of the National Academy of
  Sciences}\ }\textbf {\bibinfo {volume} {105}},\ \bibinfo {pages} {1232}
  (\bibinfo {year} {2008})}\BibitemShut {NoStop}%
\bibitem [{\citenamefont {{J. L. Silverberg, M. Bierbaum, J. P. Sethna, I.
  Cohen}}(2013)}]{Silverberg2013}%
  \BibitemOpen
  \bibfield  {author} {\bibinfo {author} {\bibnamefont {{J. L. Silverberg, M.
  Bierbaum, J. P. Sethna, I. Cohen}}},\ }\bibfield  {title} {\bibinfo {title}
  {Collective motion of humans in mosh and circle pits at heavy metal
  concerts},\ }\href {https://doi.org/10.1103/physrevlett.110.228701}
  {\bibfield  {journal} {\bibinfo  {journal} {Physical Review Letters}\
  }\textbf {\bibinfo {volume} {110}},\ \bibinfo {pages} {228701} (\bibinfo
  {year} {2013})}\BibitemShut {NoStop}%
\bibitem [{\citenamefont {{A. Garcimart\'{\i}n, J. M. Pastor, L. M. Ferrer, J.
  J. Ramos, C. Mart\'{\i}n-G\'omez, I. Zuriguel}}(2015)}]{Garcimart2015}%
  \BibitemOpen
  \bibfield  {author} {\bibinfo {author} {\bibnamefont {{A. Garcimart\'{\i}n,
  J. M. Pastor, L. M. Ferrer, J. J. Ramos, C. Mart\'{\i}n-G\'omez, I.
  Zuriguel}}},\ }\bibfield  {title} {\bibinfo {title} {Flow and clogging of a
  sheep herd passing through a bottleneck},\ }\href
  {https://doi.org/10.1103/PhysRevE.91.022808} {\bibfield  {journal} {\bibinfo
  {journal} {Phys. Rev. E}\ }\textbf {\bibinfo {volume} {91}},\ \bibinfo
  {pages} {022808} (\bibinfo {year} {2015})}\BibitemShut {NoStop}%
\bibitem [{\citenamefont {{A. Cavagna, I. Giardina}}(2014)}]{Cavagna2014}%
  \BibitemOpen
  \bibfield  {author} {\bibinfo {author} {\bibnamefont {{A. Cavagna, I.
  Giardina}}},\ }\bibfield  {title} {\bibinfo {title} {Bird flocks as condensed
  matter},\ }\href {https://doi.org/10.1146/annurev-conmatphys-031113-133834}
  {\bibfield  {journal} {\bibinfo  {journal} {Annual Review of Condensed Matter
  Physics}\ }\textbf {\bibinfo {volume} {5}},\ \bibinfo {pages} {183} (\bibinfo
  {year} {2014})}\BibitemShut {NoStop}%
\bibitem [{\citenamefont {{S. Ramaswamy}}(2010)}]{RamaswamyReview2010}%
  \BibitemOpen
  \bibfield  {author} {\bibinfo {author} {\bibnamefont {{S. Ramaswamy}}},\
  }\bibfield  {title} {\bibinfo {title} {The mechanics and statistics of active
  matter},\ }\href {https://doi.org/10.1146/annurev-conmatphys-070909-104101}
  {\bibfield  {journal} {\bibinfo  {journal} {Annual Review of Condensed Matter
  Physics}\ }\textbf {\bibinfo {volume} {1}},\ \bibinfo {pages} {323} (\bibinfo
  {year} {2010})}\BibitemShut {NoStop}%
\bibitem [{\citenamefont {{M. C. Marchetti, J. F. Joanny, S. Ramaswamy, T. B.
  Liverpool, J. Prost, Madan Rao, R. Aditi
  Simha}}(2013)}]{MarchettiRevModPhys2013}%
  \BibitemOpen
  \bibfield  {author} {\bibinfo {author} {\bibnamefont {{M. C. Marchetti, J. F.
  Joanny, S. Ramaswamy, T. B. Liverpool, J. Prost, Madan Rao, R. Aditi
  Simha}}},\ }\bibfield  {title} {\bibinfo {title} {Hydrodynamics of soft
  active matter},\ }\href {https://doi.org/10.1103/revmodphys.85.1143}
  {\bibfield  {journal} {\bibinfo  {journal} {Reviews of Modern Physics}\
  }\textbf {\bibinfo {volume} {85}},\ \bibinfo {pages} {1143} (\bibinfo {year}
  {2013})}\BibitemShut {NoStop}%
\bibitem [{\citenamefont {{C. J. Olson Reichhardt, C.
  Reichhardt}}(2017)}]{ReichhardtARCMPRev2017}%
  \BibitemOpen
  \bibfield  {author} {\bibinfo {author} {\bibnamefont {{C. J. Olson
  Reichhardt, C. Reichhardt}}},\ }\bibfield  {title} {\bibinfo {title} {Ratchet
  effects in active matter systems},\ }\href
  {https://doi.org/10.1146/annurev-conmatphys-031016-025522} {\bibfield
  {journal} {\bibinfo  {journal} {Annual Review of Condensed Matter Physics}\
  }\textbf {\bibinfo {volume} {8}},\ \bibinfo {pages} {51} (\bibinfo {year}
  {2017})}\BibitemShut {NoStop}%
\bibitem [{\citenamefont {{A. M. Menzel}}(2015)}]{Menzel2015}%
  \BibitemOpen
  \bibfield  {author} {\bibinfo {author} {\bibnamefont {{A. M. Menzel}}},\
  }\bibfield  {title} {\bibinfo {title} {Tuned, driven, and active soft
  matter},\ }\href {https://doi.org/10.1016/j.physrep.2014.10.001} {\bibfield
  {journal} {\bibinfo  {journal} {Physics Reports}\ }\textbf {\bibinfo {volume}
  {554}},\ \bibinfo {pages} {1} (\bibinfo {year} {2015})}\BibitemShut {NoStop}%
\bibitem [{\citenamefont {{C. Bechinger, R. D. Leonardo, H. L\"{o}wen, C.
  Reichhardt, G. Volpe, G. Volpe}}(2016)}]{BechingerReview2016}%
  \BibitemOpen
  \bibfield  {author} {\bibinfo {author} {\bibnamefont {{C. Bechinger, R. D.
  Leonardo, H. L\"{o}wen, C. Reichhardt, G. Volpe, G. Volpe}}},\ }\bibfield
  {title} {\bibinfo {title} {Active particles in complex and crowded
  environments},\ }\href {https://doi.org/10.1103/revmodphys.88.045006}
  {\bibfield  {journal} {\bibinfo  {journal} {Reviews of Modern Physics}\
  }\textbf {\bibinfo {volume} {88}},\ \bibinfo {pages} {045006} (\bibinfo
  {year} {2016})}\BibitemShut {NoStop}%
\bibitem [{\citenamefont {{C. Marchetti, Y. Fily, S. Henkes, A. Patch, D.
  Yllanes}}(2016)}]{MarchettiCurrOp2016}%
  \BibitemOpen
  \bibfield  {author} {\bibinfo {author} {\bibnamefont {{C. Marchetti, Y. Fily,
  S. Henkes, A. Patch, D. Yllanes}}},\ }\bibfield  {title} {\bibinfo {title}
  {Minimal model of active colloids highlights the role of mechanical
  interactions in controlling the emergent behavior of active matter},\ }\href
  {https://doi.org/10.1016/j.cocis.2016.01.003} {\bibfield  {journal} {\bibinfo
   {journal} {Current Opinion in Colloid {\&} Interface Science}\ }\textbf
  {\bibinfo {volume} {21}},\ \bibinfo {pages} {34} (\bibinfo {year}
  {2016})}\BibitemShut {NoStop}%
\bibitem [{\citenamefont {{M. Cates, J.
  Tailleur}}(2015)}]{TailleurCatesRev2015}%
  \BibitemOpen
  \bibfield  {author} {\bibinfo {author} {\bibnamefont {{M. Cates, J.
  Tailleur}}},\ }\bibfield  {title} {\bibinfo {title} {Motility-induced phase
  separation},\ }\href
  {https://doi.org/10.1146/annurev-conmatphys-031214-014710} {\bibfield
  {journal} {\bibinfo  {journal} {Annual Review of Condensed Matter Physics}\
  }\textbf {\bibinfo {volume} {6}},\ \bibinfo {pages} {219} (\bibinfo {year}
  {2015})}\BibitemShut {NoStop}%
\bibitem [{\citenamefont {{T. Vicsek, A. Zafeiris}}(2012)}]{Vicsek2012}%
  \BibitemOpen
  \bibfield  {author} {\bibinfo {author} {\bibnamefont {{T. Vicsek, A.
  Zafeiris}}},\ }\bibfield  {title} {\bibinfo {title} {Collective motion},\
  }\href {https://doi.org/10.1016/j.physrep.2012.03.004} {\bibfield  {journal}
  {\bibinfo  {journal} {Physics Reports}\ }\textbf {\bibinfo {volume} {517}},\
  \bibinfo {pages} {71} (\bibinfo {year} {2012})}\BibitemShut {NoStop}%
\bibitem [{\citenamefont {{O. Holland C.
  Melhuish}}(1999)}]{holland1999stigmergy}%
  \BibitemOpen
  \bibfield  {author} {\bibinfo {author} {\bibnamefont {{O. Holland C.
  Melhuish}}},\ }\bibfield  {title} {\bibinfo {title} {Stigmergy,
  self-organization, and sorting in collective robotics},\ }\href
  {https://doi.org/10.1162/106454699568737} {\bibfield  {journal} {\bibinfo
  {journal} {Artificial Life}\ }\textbf {\bibinfo {volume} {5}},\ \bibinfo
  {pages} {173} (\bibinfo {year} {1999})}\BibitemShut {NoStop}%
\bibitem [{\citenamefont {{H. Hamann}}(2018)}]{hamann2018swarm}%
  \BibitemOpen
  \bibfield  {author} {\bibinfo {author} {\bibnamefont {{H. Hamann}}},\ }\href
  {https://doi.org/10.1007/978-3-319-74528-2} {\emph {\bibinfo {title} {Swarm
  Robotics: A Formal Approach}}}\ (\bibinfo  {publisher} {Springer
  International Publishing},\ \bibinfo {year} {2018})\BibitemShut {NoStop}%
\bibitem [{\citenamefont {{Y. Tan}}(2016)}]{tan2015handbook}%
  \BibitemOpen
  \bibinfo {editor} {\bibnamefont {{Y. Tan}}},\ ed.,\ \href
  {https://doi.org/10.4018/978-1-4666-9572-6} {\emph {\bibinfo {title}
  {Handbook of Research on Design, Control, and Modeling of Swarm Robotics}}}\
  (\bibinfo  {publisher} {{IGI} Global},\ \bibinfo {year} {2016})\BibitemShut
  {NoStop}%
\bibitem [{\citenamefont {{S. Kernbach}}(2013)}]{kernbach2013handbook}%
  \BibitemOpen
  \bibinfo {editor} {\bibnamefont {{S. Kernbach}}},\ ed.,\ \href
  {https://doi.org/10.1201/b14908} {\emph {\bibinfo {title} {Handbook of
  Collective Robotics}}}\ (\bibinfo  {publisher} {Jenny Stanford Publishing},\
  \bibinfo {year} {2013})\BibitemShut {NoStop}%
\bibitem [{\citenamefont {{M. Rubenstein, A. Cornejo, R.
  Nagpal}}(2014)}]{Rubenstein2014}%
  \BibitemOpen
  \bibfield  {author} {\bibinfo {author} {\bibnamefont {{M. Rubenstein, A.
  Cornejo, R. Nagpal}}},\ }\bibfield  {title} {\bibinfo {title} {Programmable
  self-assembly in a thousand-robot swarm},\ }\href
  {https://doi.org/10.1126/science.1254295} {\bibfield  {journal} {\bibinfo
  {journal} {Science}\ }\textbf {\bibinfo {volume} {345}},\ \bibinfo {pages}
  {795} (\bibinfo {year} {2014})}\BibitemShut {NoStop}%
\bibitem [{\citenamefont {{R. Gro, M. Bonani, F. Mondada, M.
  Dorigo}}(2006)}]{Gro2006}%
  \BibitemOpen
  \bibfield  {author} {\bibinfo {author} {\bibnamefont {{R. Gro, M. Bonani, F.
  Mondada, M. Dorigo}}},\ }\bibfield  {title} {\bibinfo {title} {Autonomous
  self-assembly in swarm-bots},\ }\href
  {https://doi.org/10.1109/tro.2006.882919} {\bibfield  {journal} {\bibinfo
  {journal} {{IEEE} Transactions on Robotics}\ }\textbf {\bibinfo {volume}
  {22}},\ \bibinfo {pages} {1115} (\bibinfo {year} {2006})}\BibitemShut
  {NoStop}%
\bibitem [{\citenamefont {{M. Yim, W. Shen, B. Salemi, D. Rus, M. Moll, H.
  Lipson, E. Klavins, G. Chirikjian}}(2007)}]{Yim2007}%
  \BibitemOpen
  \bibfield  {author} {\bibinfo {author} {\bibnamefont {{M. Yim, W. Shen, B.
  Salemi, D. Rus, M. Moll, H. Lipson, E. Klavins, G. Chirikjian}}},\ }\bibfield
   {title} {\bibinfo {title} {Modular self-reconfigurable robot systems [grand
  challenges of robotics]},\ }\href {https://doi.org/10.1109/mra.2007.339623}
  {\bibfield  {journal} {\bibinfo  {journal} {{IEEE} Robotics {\&} Automation
  Magazine}\ }\textbf {\bibinfo {volume} {14}},\ \bibinfo {pages} {43}
  (\bibinfo {year} {2007})}\BibitemShut {NoStop}%
\bibitem [{\citenamefont {{S.C. Goldstein, J.D. Campbell, T.C.
  Mowry}}(2005)}]{Goldstein2005}%
  \BibitemOpen
  \bibfield  {author} {\bibinfo {author} {\bibnamefont {{S.C. Goldstein, J.D.
  Campbell, T.C. Mowry}}},\ }\bibfield  {title} {\bibinfo {title} {Programmable
  matter},\ }\href {https://doi.org/10.1109/mc.2005.198} {\bibfield  {journal}
  {\bibinfo  {journal} {Computer}\ }\textbf {\bibinfo {volume} {38}},\ \bibinfo
  {pages} {99} (\bibinfo {year} {2005})}\BibitemShut {NoStop}%
\bibitem [{\citenamefont {{Vaughan, Adam Lein1 Richard
  T}}(2008)}]{vaughan2008adaptive}%
  \BibitemOpen
  \bibfield  {author} {\bibinfo {author} {\bibnamefont {{Vaughan, Adam Lein1
  Richard T}}},\ }\bibfield  {title} {\bibinfo {title} {Adaptive multi-robot
  bucket brigade foraging},\ }\href@noop {} {\bibfield  {journal} {\bibinfo
  {journal} {Artificial Life}\ }\textbf {\bibinfo {volume} {11}},\ \bibinfo
  {pages} {337} (\bibinfo {year} {2008})}\BibitemShut {NoStop}%
\bibitem [{\citenamefont {{L. E. Parker, D. Rus, G. S.
  Sukhatme}}(2016)}]{parker2008multiple}%
  \BibitemOpen
  \bibfield  {author} {\bibinfo {author} {\bibnamefont {{L. E. Parker, D. Rus,
  G. S. Sukhatme}}},\ }\bibfield  {title} {\bibinfo {title} {Multiple mobile
  robot systems},\ }in\ \href {https://doi.org/10.1007/978-3-319-32552-1_53}
  {\emph {\bibinfo {booktitle} {Springer Handbook of Robotics}}}\ (\bibinfo
  {publisher} {Springer International Publishing},\ \bibinfo {year} {2016})\
  pp.\ \bibinfo {pages} {1335--1384}\BibitemShut {NoStop}%
\bibitem [{\citenamefont {{K. Konolige, D. Fox, C. Ortiz, A. Agno, M. Eriksen,
  B. Limketkai, J. Ko, B. Morisset, D. Schulz, B. Stewart, R.
  Vincent}}(2006)}]{Konolige2006}%
  \BibitemOpen
  \bibfield  {author} {\bibinfo {author} {\bibnamefont {{K. Konolige, D. Fox,
  C. Ortiz, A. Agno, M. Eriksen, B. Limketkai, J. Ko, B. Morisset, D. Schulz,
  B. Stewart, R. Vincent}}},\ }\bibfield  {title} {\bibinfo {title} {Centibots:
  Very large scale distributed robotic teams},\ }in\ \href
  {https://doi.org/10.1007/11552246_13} {\emph {\bibinfo {booktitle} {Springer
  Tracts in Advanced Robotics}}}\ (\bibinfo  {publisher} {Springer Berlin
  Heidelberg},\ \bibinfo {year} {2006})\ pp.\ \bibinfo {pages}
  {131--140}\BibitemShut {NoStop}%
\bibitem [{\citenamefont {{H. H. Wensink, H. L\"owen}}(2008)}]{Wensink2008}%
  \BibitemOpen
  \bibfield  {author} {\bibinfo {author} {\bibnamefont {{H. H. Wensink, H.
  L\"owen}}},\ }\bibfield  {title} {\bibinfo {title} {Aggregation of
  self-propelled colloidal rods near confining walls},\ }\href
  {https://doi.org/10.1103/PhysRevE.78.031409} {\bibfield  {journal} {\bibinfo
  {journal} {Phys. Rev. E}\ }\textbf {\bibinfo {volume} {78}},\ \bibinfo
  {pages} {031409} (\bibinfo {year} {2008})}\BibitemShut {NoStop}%
\bibitem [{\citenamefont {{X. Yang, L. M. Manning, C. M.
  Marchetti}}(2014)}]{Yang2014}%
  \BibitemOpen
  \bibfield  {author} {\bibinfo {author} {\bibnamefont {{X. Yang, L. M.
  Manning, C. M. Marchetti}}},\ }\bibfield  {title} {\bibinfo {title}
  {Aggregation and segregation of confined active particles},\ }\href
  {https://doi.org/10.1039/c4sm00927d} {\bibfield  {journal} {\bibinfo
  {journal} {Soft Matter}\ }\textbf {\bibinfo {volume} {10}},\ \bibinfo {pages}
  {6477} (\bibinfo {year} {2014})}\BibitemShut {NoStop}%
\bibitem [{\citenamefont {{C. F. Lee}}(2013)}]{Lee2013}%
  \BibitemOpen
  \bibfield  {author} {\bibinfo {author} {\bibnamefont {{C. F. Lee}}},\
  }\bibfield  {title} {\bibinfo {title} {Active particles under confinement:
  aggregation at the wall and gradient formation inside a channel},\ }\href
  {https://doi.org/10.1088/1367-2630/15/5/055007} {\bibfield  {journal}
  {\bibinfo  {journal} {New Journal of Physics}\ }\textbf {\bibinfo {volume}
  {15}},\ \bibinfo {pages} {055007} (\bibinfo {year} {2013})}\BibitemShut
  {NoStop}%
\bibitem [{\citenamefont {{J. Elgeti, G. Gompper}}(2009)}]{Elgeti2009}%
  \BibitemOpen
  \bibfield  {author} {\bibinfo {author} {\bibnamefont {{J. Elgeti, G.
  Gompper}}},\ }\bibfield  {title} {\bibinfo {title} {Self-propelled rods near
  surfaces},\ }\href {https://doi.org/10.1209/0295-5075/85/38002} {\bibfield
  {journal} {\bibinfo  {journal} {{EPL} (Europhysics Letters)}\ }\textbf
  {\bibinfo {volume} {85}},\ \bibinfo {pages} {38002} (\bibinfo {year}
  {2009})}\BibitemShut {NoStop}%
\bibitem [{\citenamefont {{J. Elgeti, G. Gompper}}(2013)}]{Elgeti2013}%
  \BibitemOpen
  \bibfield  {author} {\bibinfo {author} {\bibnamefont {{J. Elgeti, G.
  Gompper}}},\ }\bibfield  {title} {\bibinfo {title} {Wall accumulation of
  self-propelled spheres},\ }\href
  {https://doi.org/10.1209/0295-5075/101/48003} {\bibfield  {journal} {\bibinfo
   {journal} {{EPL} (Europhysics Letters)}\ }\textbf {\bibinfo {volume}
  {101}},\ \bibinfo {pages} {48003} (\bibinfo {year} {2013})}\BibitemShut
  {NoStop}%
\bibitem [{\citenamefont {{L. Caprini, U. M. B. Marconi}}(2018)}]{caprini2018}%
  \BibitemOpen
  \bibfield  {author} {\bibinfo {author} {\bibnamefont {{L. Caprini, U. M. B.
  Marconi}}},\ }\bibfield  {title} {\bibinfo {title} {Active particles under
  confinement and effective force generation among surfaces},\ }\href
  {https://doi.org/10.1039/c8sm01840e} {\bibfield  {journal} {\bibinfo
  {journal} {Soft Matter}\ }\textbf {\bibinfo {volume} {14}},\ \bibinfo {pages}
  {9044} (\bibinfo {year} {2018})}\BibitemShut {NoStop}%
\bibitem [{\citenamefont {{Y. Fily, A. Baskaran, M. F.
  Hagan}}(2015)}]{Fily2015}%
  \BibitemOpen
  \bibfield  {author} {\bibinfo {author} {\bibnamefont {{Y. Fily, A. Baskaran,
  M. F. Hagan}}},\ }\bibfield  {title} {\bibinfo {title} {Dynamics and density
  distribution of strongly confined noninteracting nonaligning self-propelled
  particles in a nonconvex boundary},\ }\href
  {https://doi.org/10.1103/physreve.91.012125} {\bibfield  {journal} {\bibinfo
  {journal} {Physical Review E}\ }\textbf {\bibinfo {volume} {91}},\ \bibinfo
  {pages} {012125} (\bibinfo {year} {2015})}\BibitemShut {NoStop}%
\bibitem [{\citenamefont {{T. Hiraoka, T. Shimada, N.
  Ito}}(2017)}]{Hiraoka2017}%
  \BibitemOpen
  \bibfield  {author} {\bibinfo {author} {\bibnamefont {{T. Hiraoka, T.
  Shimada, N. Ito}}},\ }\bibfield  {title} {\bibinfo {title} {Collective motion
  in repulsive self-propelled particles in confined geometries},\ }\href
  {https://doi.org/10.1088/1742-6596/921/1/012006} {\bibfield  {journal}
  {\bibinfo  {journal} {Journal of Physics: Conference Series}\ }\textbf
  {\bibinfo {volume} {921}},\ \bibinfo {pages} {012006} (\bibinfo {year}
  {2017})}\BibitemShut {NoStop}%
\bibitem [{\citenamefont {{H. Levine, W. Rappel, I.
  Cohen}}(2000)}]{Levine2000}%
  \BibitemOpen
  \bibfield  {author} {\bibinfo {author} {\bibnamefont {{H. Levine, W. Rappel,
  I. Cohen}}},\ }\bibfield  {title} {\bibinfo {title} {Self-organization in
  systems of self-propelled particles},\ }\href
  {https://doi.org/10.1103/PhysRevE.63.017101} {\bibfield  {journal} {\bibinfo
  {journal} {Phys. Rev. E}\ }\textbf {\bibinfo {volume} {63}},\ \bibinfo
  {pages} {017101} (\bibinfo {year} {2000})}\BibitemShut {NoStop}%
\bibitem [{\citenamefont {{M. S. E. Peterson, A. Baskaran, M. F.
  Hagan}}(2021)}]{Peterson2021}%
  \BibitemOpen
  \bibfield  {author} {\bibinfo {author} {\bibnamefont {{M. S. E. Peterson, A.
  Baskaran, M. F. Hagan}}},\ }\bibfield  {title} {\bibinfo {title} {Vesicle
  shape transformations driven by confined active filaments},\ }\href
  {https://doi.org/10.1038/s41467-021-27310-8} {\bibfield  {journal} {\bibinfo
  {journal} {Nature Communications}\ }\textbf {\bibinfo {volume} {12}},\
  \bibinfo {pages} {7247} (\bibinfo {year} {2021})}\BibitemShut {NoStop}%
\bibitem [{\citenamefont {{N. Nikola, A. P. Solon, Y. Kafri, M. Kardar, J.
  Tailleur, R. Voituriez}}(2016)}]{Nikola2016}%
  \BibitemOpen
  \bibfield  {author} {\bibinfo {author} {\bibnamefont {{N. Nikola, A. P.
  Solon, Y. Kafri, M. Kardar, J. Tailleur, R. Voituriez}}},\ }\bibfield
  {title} {\bibinfo {title} {Active particles with soft and curved walls:
  Equation of state, ratchets, and instabilities},\ }\href
  {https://doi.org/10.1103/physrevlett.117.098001} {\bibfield  {journal}
  {\bibinfo  {journal} {Physical Review Letters}\ }\textbf {\bibinfo {volume}
  {117}},\ \bibinfo {pages} {098001} (\bibinfo {year} {2016})}\BibitemShut
  {NoStop}%
\bibitem [{\citenamefont {{M. Paoluzzi, R. D. Leonardo, C. M. Marchetti, L.
  Angelani}}(2016)}]{Paoluzzi2016}%
  \BibitemOpen
  \bibfield  {author} {\bibinfo {author} {\bibnamefont {{M. Paoluzzi, R. D.
  Leonardo, C. M. Marchetti, L. Angelani}}},\ }\bibfield  {title} {\bibinfo
  {title} {Shape and displacement fluctuations in soft vesicles filled by
  active particles},\ }\href {https://doi.org/10.1038/srep34146} {\bibfield
  {journal} {\bibinfo  {journal} {Scientific Reports}\ }\textbf {\bibinfo
  {volume} {6}},\ \bibinfo {pages} {34146} (\bibinfo {year}
  {2016})}\BibitemShut {NoStop}%
\bibitem [{\citenamefont {{W. Tian, Y. Gu, Y. Guo, K.
  Chen}}(2017)}]{Tian_2017}%
  \BibitemOpen
  \bibfield  {author} {\bibinfo {author} {\bibnamefont {{W. Tian, Y. Gu, Y.
  Guo, K. Chen}}},\ }\bibfield  {title} {\bibinfo {title} {Anomalous boundary
  deformation induced by enclosed active particles},\ }\href
  {https://doi.org/10.1088/1674-1056/26/10/100502} {\bibfield  {journal}
  {\bibinfo  {journal} {Chinese Physics B}\ }\textbf {\bibinfo {volume} {26}},\
  \bibinfo {pages} {100502} (\bibinfo {year} {2017})}\BibitemShut {NoStop}%
\bibitem [{\citenamefont {{Y. Li, P. R. ten Wolde}}(2019)}]{Li2019}%
  \BibitemOpen
  \bibfield  {author} {\bibinfo {author} {\bibnamefont {{Y. Li, P. R. ten
  Wolde}}},\ }\bibfield  {title} {\bibinfo {title} {Shape transformations of
  vesicles induced by swim pressure},\ }\href
  {https://doi.org/10.1103/physrevlett.123.148003} {\bibfield  {journal}
  {\bibinfo  {journal} {Physical Review Letters}\ }\textbf {\bibinfo {volume}
  {123}},\ \bibinfo {pages} {148003} (\bibinfo {year} {2019})}\BibitemShut
  {NoStop}%
\bibitem [{\citenamefont {{C. Wang, Y. Guo, W. Tian, K.
  Chen}}(2019)}]{Wang2019}%
  \BibitemOpen
  \bibfield  {author} {\bibinfo {author} {\bibnamefont {{C. Wang, Y. Guo, W.
  Tian, K. Chen}}},\ }\bibfield  {title} {\bibinfo {title} {Shape
  transformation and manipulation of a vesicle by active particles},\ }\href
  {https://doi.org/10.1063/1.5078694} {\bibfield  {journal} {\bibinfo
  {journal} {The Journal of Chemical Physics}\ }\textbf {\bibinfo {volume}
  {150}},\ \bibinfo {pages} {044907} (\bibinfo {year} {2019})}\BibitemShut
  {NoStop}%
\bibitem [{\citenamefont {{A. Deblais, T. Barois, T. Guerin,
  P.{\hspace{0.167em}}H. Delville, R. Vaudaine, J.{\hspace{0.167em}}S.
  Lintuvuori, J.{\hspace{0.167em}}F. Boudet, J.{\hspace{0.167em}}C. Baret, H.
  Kellay}}(2018)}]{Deblais2018}%
  \BibitemOpen
  \bibfield  {author} {\bibinfo {author} {\bibnamefont {{A. Deblais, T. Barois,
  T. Guerin, P.{\hspace{0.167em}}H. Delville, R. Vaudaine,
  J.{\hspace{0.167em}}S. Lintuvuori, J.{\hspace{0.167em}}F. Boudet,
  J.{\hspace{0.167em}}C. Baret, H. Kellay}}},\ }\bibfield  {title} {\bibinfo
  {title} {Boundaries control collective dynamics of inertial self-propelled
  robots},\ }\href {https://doi.org/10.1103/physrevlett.120.188002} {\bibfield
  {journal} {\bibinfo  {journal} {Physical Review Letters}\ }\textbf {\bibinfo
  {volume} {120}},\ \bibinfo {pages} {188002} (\bibinfo {year}
  {2018})}\BibitemShut {NoStop}%
\bibitem [{\citenamefont {{O. Feinerman, I. Pinkoviezky, A. Gelblum, E. Fonio,
  N. S. Gov}}(2018)}]{Feinerman2018}%
  \BibitemOpen
  \bibfield  {author} {\bibinfo {author} {\bibnamefont {{O. Feinerman, I.
  Pinkoviezky, A. Gelblum, E. Fonio, N. S. Gov}}},\ }\bibfield  {title}
  {\bibinfo {title} {The physics of cooperative transport in groups of ants},\
  }\href {https://doi.org/10.1038/s41567-018-0107-y} {\bibfield  {journal}
  {\bibinfo  {journal} {Nature Physics}\ }\textbf {\bibinfo {volume} {14}},\
  \bibinfo {pages} {683} (\bibinfo {year} {2018})}\BibitemShut {NoStop}%
\bibitem [{\citenamefont {{S. Reuveni}}(2016)}]{Reuveni2016}%
  \BibitemOpen
  \bibfield  {author} {\bibinfo {author} {\bibnamefont {{S. Reuveni}}},\
  }\bibfield  {title} {\bibinfo {title} {Optimal stochastic restart renders
  fluctuations in first passage times universal},\ }\href
  {https://doi.org/10.1103/physrevlett.116.170601} {\bibfield  {journal}
  {\bibinfo  {journal} {Physical Review Letters}\ }\textbf {\bibinfo {volume}
  {116}},\ \bibinfo {pages} {170601} (\bibinfo {year} {2016})}\BibitemShut
  {NoStop}%
\bibitem [{\citenamefont {{R. E. Isele-Holder, J. Elgeti, G.
  Gompper}}(2015)}]{Isele-Holder2015}%
  \BibitemOpen
  \bibfield  {author} {\bibinfo {author} {\bibnamefont {{R. E. Isele-Holder, J.
  Elgeti, G. Gompper}}},\ }\bibfield  {title} {\bibinfo {title} {Self-propelled
  worm-like filaments: spontaneous spiral formation, structure, and dynamics},\
  }\href {https://doi.org/10.1039/c5sm01683e} {\bibfield  {journal} {\bibinfo
  {journal} {Soft Matter}\ }\textbf {\bibinfo {volume} {11}},\ \bibinfo {pages}
  {7181} (\bibinfo {year} {2015})}\BibitemShut {NoStop}%
\bibitem [{\citenamefont {{R. E. Isele-Holder, J. J\"{a}ger, G. Saggiorato, J.
  Elgeti, G. Gompper}}(2016)}]{Isele-Holder2016}%
  \BibitemOpen
  \bibfield  {author} {\bibinfo {author} {\bibnamefont {{R. E. Isele-Holder, J.
  J\"{a}ger, G. Saggiorato, J. Elgeti, G. Gompper}}},\ }\bibfield  {title}
  {\bibinfo {title} {Dynamics of self-propelled filaments pushing a load},\
  }\href {https://doi.org/10.1039/c6sm01094f} {\bibfield  {journal} {\bibinfo
  {journal} {Soft Matter}\ }\textbf {\bibinfo {volume} {12}},\ \bibinfo {pages}
  {8495} (\bibinfo {year} {2016})}\BibitemShut {NoStop}%
\bibitem [{\citenamefont {{\"{O}. Duman, R. E. Isele-Holder, J. Elgeti, G.
  Gompper}}(2018)}]{Duman2018}%
  \BibitemOpen
  \bibfield  {author} {\bibinfo {author} {\bibnamefont {{\"{O}. Duman, R. E.
  Isele-Holder, J. Elgeti, G. Gompper}}},\ }\bibfield  {title} {\bibinfo
  {title} {Collective dynamics of self-propelled semiflexible filaments},\
  }\href {https://doi.org/10.1039/c8sm00282g} {\bibfield  {journal} {\bibinfo
  {journal} {Soft Matter}\ }\textbf {\bibinfo {volume} {14}},\ \bibinfo {pages}
  {4483} (\bibinfo {year} {2018})}\BibitemShut {NoStop}%
\bibitem [{\citenamefont {Chelakkot}\ \emph {et~al.}(2021)\citenamefont
  {Chelakkot}, \citenamefont {Hagan},\ and\ \citenamefont
  {Gopinath}}]{Chelakkot2020}%
  \BibitemOpen
  \bibfield  {author} {\bibinfo {author} {\bibfnamefont {R.}~\bibnamefont
  {Chelakkot}}, \bibinfo {author} {\bibfnamefont {M.~F.}\ \bibnamefont
  {Hagan}},\ and\ \bibinfo {author} {\bibfnamefont {A.}~\bibnamefont
  {Gopinath}},\ }\bibfield  {title} {\bibinfo {title} {Synchronized
  oscillations, traveling waves, and jammed clusters induced by steric
  interactions in active filament arrays},\ }\href
  {https://doi.org/10.1039/d0sm01162b} {\bibfield  {journal} {\bibinfo
  {journal} {Soft Matter}\ }\textbf {\bibinfo {volume} {17}},\ \bibinfo {pages}
  {1091} (\bibinfo {year} {2021})}\BibitemShut {NoStop}%
\bibitem [{\citenamefont {{J. D. Weeks, D. Chandler, H. C.
  Andersen}}(1971)}]{Weeks1971}%
  \BibitemOpen
  \bibfield  {author} {\bibinfo {author} {\bibnamefont {{J. D. Weeks, D.
  Chandler, H. C. Andersen}}},\ }\bibfield  {title} {\bibinfo {title} {Role of
  repulsive forces in determining the equilibrium structure of simple
  liquids},\ }\href {https://doi.org/10.1063/1.1674820} {\bibfield  {journal}
  {\bibinfo  {journal} {The Journal of Chemical Physics}\ }\textbf {\bibinfo
  {volume} {54}},\ \bibinfo {pages} {5237} (\bibinfo {year}
  {1971})}\BibitemShut {NoStop}%
\bibitem [{\citenamefont {{H. C. Andersen, J. D. Weeks, D.
  Chandler}}(1971)}]{Andersen1971}%
  \BibitemOpen
  \bibfield  {author} {\bibinfo {author} {\bibnamefont {{H. C. Andersen, J. D.
  Weeks, D. Chandler}}},\ }\bibfield  {title} {\bibinfo {title} {Relationship
  between the hard-sphere fluid and fluids with realistic repulsive forces},\
  }\href {https://doi.org/10.1103/physreva.4.1597} {\bibfield  {journal}
  {\bibinfo  {journal} {Physical Review A}\ }\textbf {\bibinfo {volume} {4}},\
  \bibinfo {pages} {1597} (\bibinfo {year} {1971})}\BibitemShut {NoStop}%
\bibitem [{\citenamefont {{Y. Fily, A. Baskaran, M. F.
  Hagan}}(2014)}]{Fily2014}%
  \BibitemOpen
  \bibfield  {author} {\bibinfo {author} {\bibnamefont {{Y. Fily, A. Baskaran,
  M. F. Hagan}}},\ }\bibfield  {title} {\bibinfo {title} {Dynamics of
  self-propelled particles under strong confinement},\ }\href
  {https://doi.org/10.1039/c4sm00975d} {\bibfield  {journal} {\bibinfo
  {journal} {Soft Matter}\ }\textbf {\bibinfo {volume} {10}},\ \bibinfo {pages}
  {5609} (\bibinfo {year} {2014})}\BibitemShut {NoStop}%
\bibitem [{\citenamefont {{J. Tian, Y. Liu, J. Chen, B. Guo, S.
  Prasad}}(2021)}]{Tian2021}%
  \BibitemOpen
  \bibfield  {author} {\bibinfo {author} {\bibnamefont {{J. Tian, Y. Liu, J.
  Chen, B. Guo, S. Prasad}}},\ }\bibfield  {title} {\bibinfo {title} {Finite
  element analysis of a self-propelled capsule robot moving in the small
  intestine},\ }\href {https://doi.org/10.1016/j.ijmecsci.2021.106621}
  {\bibfield  {journal} {\bibinfo  {journal} {International Journal of
  Mechanical Sciences}\ }\textbf {\bibinfo {volume} {206}},\ \bibinfo {pages}
  {106621} (\bibinfo {year} {2021})}\BibitemShut {NoStop}%
\bibitem [{\citenamefont {{F. Smallenburg, H.
  L\"{o}wen}}(2015)}]{Smallenburg2015}%
  \BibitemOpen
  \bibfield  {author} {\bibinfo {author} {\bibnamefont {{F. Smallenburg, H.
  L\"{o}wen}}},\ }\bibfield  {title} {\bibinfo {title} {Swim pressure on walls
  with curves and corners},\ }\href
  {https://doi.org/10.1103/physreve.92.032304} {\bibfield  {journal} {\bibinfo
  {journal} {Physical Review E}\ }\textbf {\bibinfo {volume} {92}},\ \bibinfo
  {pages} {032304} (\bibinfo {year} {2015})}\BibitemShut {NoStop}%
\bibitem [{\citenamefont {{J. Elgeti, G. Gompper}}(2015)}]{Elgeti2015}%
  \BibitemOpen
  \bibfield  {author} {\bibinfo {author} {\bibnamefont {{J. Elgeti, G.
  Gompper}}},\ }\bibfield  {title} {\bibinfo {title} {Run-and-tumble dynamics
  of self-propelled particles in confinement},\ }\href
  {https://doi.org/10.1209/0295-5075/109/58003} {\bibfield  {journal} {\bibinfo
   {journal} {{EPL} (Europhysics Letters)}\ }\textbf {\bibinfo {volume}
  {109}},\ \bibinfo {pages} {58003} (\bibinfo {year} {2015})}\BibitemShut
  {NoStop}%
\bibitem [{\citenamefont {{S. A. Mallory, C. Valeriani, A.
  Cacciuto}}(2014)}]{Mallory2014}%
  \BibitemOpen
  \bibfield  {author} {\bibinfo {author} {\bibnamefont {{S. A. Mallory, C.
  Valeriani, A. Cacciuto}}},\ }\bibfield  {title} {\bibinfo {title}
  {Curvature-induced activation of a passive tracer in an active bath},\ }\href
  {https://doi.org/10.1103/PhysRevE.90.032309} {\bibfield  {journal} {\bibinfo
  {journal} {Phys. Rev. E}\ }\textbf {\bibinfo {volume} {90}},\ \bibinfo
  {pages} {032309} (\bibinfo {year} {2014})}\BibitemShut {NoStop}%
\bibitem [{\citenamefont {{H. R. Vutukuri, M. Hoore, C. Abaurrea-Velasco, L.
  van Buren, A. Dutto, T. Auth, D. A. Fedosov, G. Gompper, J.
  Vermant}}(2019)}]{Vutukuri2019}%
  \BibitemOpen
  \bibfield  {author} {\bibinfo {author} {\bibnamefont {{H. R. Vutukuri, M.
  Hoore, C. Abaurrea-Velasco, L. van Buren, A. Dutto, T. Auth, D. A. Fedosov,
  G. Gompper, J. Vermant}}},\ }\href@noop {} {\bibinfo {title} {Sculpting
  vesicles with active particles: Less is more}} (\bibinfo {year} {2019}),\
  \Eprint {https://arxiv.org/abs/1911.02381} {arXiv:1911.02381 [cond-mat.soft]}
  \BibitemShut {NoStop}%
\bibitem [{\citenamefont {{M. B\"{a}r, R. Gro{\ss}mann, S. Heidenreich, F.
  Peruani}}(2020)}]{Br2020}%
  \BibitemOpen
  \bibfield  {author} {\bibinfo {author} {\bibnamefont {{M. B\"{a}r, R.
  Gro{\ss}mann, S. Heidenreich, F. Peruani}}},\ }\bibfield  {title} {\bibinfo
  {title} {Self-propelled rods: Insights and perspectives for active matter},\
  }\href {https://doi.org/10.1146/annurev-conmatphys-031119-050611} {\bibfield
  {journal} {\bibinfo  {journal} {Annual Review of Condensed Matter Physics}\
  }\textbf {\bibinfo {volume} {11}},\ \bibinfo {pages} {441} (\bibinfo {year}
  {2020})}\BibitemShut {NoStop}%
\bibitem [{\citenamefont {{W. Helfrich}}(1973)}]{Helfrich1973}%
  \BibitemOpen
  \bibfield  {author} {\bibinfo {author} {\bibnamefont {{W. Helfrich}}},\
  }\bibfield  {title} {\bibinfo {title} {Elastic properties of lipid bilayers:
  Theory and possible experiments},\ }\href
  {https://doi.org/10.1515/znc-1973-11-1209} {\bibfield  {journal} {\bibinfo
  {journal} {Zeitschrift f\"{u}r Naturforschung C}\ }\textbf {\bibinfo {volume}
  {28}},\ \bibinfo {pages} {693} (\bibinfo {year} {1973})}\BibitemShut
  {NoStop}%
\bibitem [{\citenamefont {{M. Deserno}}(2015)}]{Deserno2015}%
  \BibitemOpen
  \bibfield  {author} {\bibinfo {author} {\bibnamefont {{M. Deserno}}},\
  }\bibfield  {title} {\bibinfo {title} {Fluid lipid membranes: From
  differential geometry to curvature stresses},\ }\href
  {https://doi.org/10.1016/j.chemphyslip.2014.05.001} {\bibfield  {journal}
  {\bibinfo  {journal} {Chemistry and Physics of Lipids}\ }\textbf {\bibinfo
  {volume} {185}},\ \bibinfo {pages} {11} (\bibinfo {year} {2015})}\BibitemShut
  {NoStop}%
\bibitem [{\citenamefont {{A. {\v{S}}ari{\'{c}}, A.
  Cacciuto}}(2013)}]{Saric2013}%
  \BibitemOpen
  \bibfield  {author} {\bibinfo {author} {\bibnamefont {{A. {\v{S}}ari{\'{c}},
  A. Cacciuto}}},\ }\bibfield  {title} {\bibinfo {title} {Self-assembly of
  nanoparticles adsorbed on fluid and elastic membranes},\ }\href
  {https://doi.org/10.1039/c3sm50188d} {\bibfield  {journal} {\bibinfo
  {journal} {Soft Matter}\ }\textbf {\bibinfo {volume} {9}},\ \bibinfo {pages}
  {6677} (\bibinfo {year} {2013})}\BibitemShut {NoStop}%
\bibitem [{\citenamefont {{F. Peruani, T. Klauss, A. Deutsch, A.
  Voss-Boehme}}(2011)}]{Peruani2011}%
  \BibitemOpen
  \bibfield  {author} {\bibinfo {author} {\bibnamefont {{F. Peruani, T. Klauss,
  A. Deutsch, A. Voss-Boehme}}},\ }\bibfield  {title} {\bibinfo {title}
  {Traffic jams, gliders, and bands in the quest for collective motion of
  self-propelled particles},\ }\href
  {https://doi.org/10.1103/physrevlett.106.128101} {\bibfield  {journal}
  {\bibinfo  {journal} {Physical Review Letters}\ }\textbf {\bibinfo {volume}
  {106}},\ \bibinfo {pages} {128101} (\bibinfo {year} {2011})}\BibitemShut
  {NoStop}%
\bibitem [{\citenamefont {{S. Weitz, A. Deutsch, F.
  Peruani}}(2015)}]{Weitz2015}%
  \BibitemOpen
  \bibfield  {author} {\bibinfo {author} {\bibnamefont {{S. Weitz, A. Deutsch,
  F. Peruani}}},\ }\bibfield  {title} {\bibinfo {title} {Self-propelled rods
  exhibit a phase-separated state characterized by the presence of active
  stresses and the ejection of polar clusters},\ }\href
  {https://doi.org/10.1103/physreve.92.012322} {\bibfield  {journal} {\bibinfo
  {journal} {Physical Review E}\ }\textbf {\bibinfo {volume} {92}},\ \bibinfo
  {pages} {012322} (\bibinfo {year} {2015})}\BibitemShut {NoStop}%
\bibitem [{\citenamefont {{M. N. van der Linden, L. C. Alexander, D.
  G.{\hspace{0.167em}}A.{\hspace{0.167em}}L. Aarts, O.
  Dauchot}}(2019)}]{vanderLinden2019}%
  \BibitemOpen
  \bibfield  {author} {\bibinfo {author} {\bibnamefont {{M. N. van der Linden,
  L. C. Alexander, D. G.{\hspace{0.167em}}A.{\hspace{0.167em}}L. Aarts, O.
  Dauchot}}},\ }\bibfield  {title} {\bibinfo {title} {Interrupted motility
  induced phase separation in aligning active colloids},\ }\href
  {https://doi.org/10.1103/physrevlett.123.098001} {\bibfield  {journal}
  {\bibinfo  {journal} {Physical Review Letters}\ }\textbf {\bibinfo {volume}
  {123}},\ \bibinfo {pages} {098001} (\bibinfo {year} {2019})}\BibitemShut
  {NoStop}%
\end{thebibliography}%

\end{document}